\documentclass[twocolumn,superscriptaddress,amsmath,amssymb,floatfix,aps]{revtex4-1}
\usepackage{amsmath}    % need for subequations

\usepackage{graphicx}   % need for figures
\usepackage{verbatim}   % useful for program listings
\usepackage{color}      % use if color is used in text
\usepackage{subfigure}  % use for side-by-side figures
\usepackage{hyperref}   % use for hypertext links, including those to external documents and URLs

\usepackage{amsmath}    % need for subequations
\usepackage{graphicx}   % need for figures
\usepackage{verbatim}   % useful for program listings
\usepackage{color}      % use if color is used in text
\usepackage{subfigure}  % use for side-by-side figures
\usepackage{hyphenat}
\usepackage[normalem]{ulem}
\usepackage{array}
\usepackage{booktabs}
\usepackage{titlesec}
\usepackage{lettrine}
\usepackage{mathtools}
\usepackage{tabularx}
\usepackage{chngcntr}
\newcolumntype{G}{X}
\newcolumntype{P}{>{\hsize=.5\hsize}X}

\titleformat{\section}
  {\normalfont\scshape}{\thesection}{1em}{}

\usepackage{color}      % use if color is used in text
\usepackage{soul}

\definecolor{darkgreen}{rgb}{0.0, 0.5, 0.0}

\newcommand{\datA}{\textit{datA} }

\begin{document}
\bibliographystyle{unsrt} % Choose Phys. Rev. style for bibliographyprsty

% \title{Replication initiation in \textit{E. coli} is regulated via an origin-density sensor generating adder correlations 
% }
\title{Synchronous replication initiation of multiple origins
}
\author{Mareike Berger} \affiliation{Biochemical Networks Group, Department of Information in Matter, AMOLF, 1098 XG Amsterdam, The Netherlands}
 \author{Pieter
  Rein ten Wolde} \affiliation{Biochemical Networks Group, Department of Information in Matter, AMOLF, 1098 XG Amsterdam, The Netherlands}

\begin{abstract} 
%Many bacterial species can contain more than one copy of their chromosome. 
Initiating replication synchronously at multiple origins of replication allows the bacterium \textit{Escherichia coli} to divide even faster than the time it takes to replicate the entire chromosome in nutrient-rich environments. What mechanisms give rise to synchronous replication initiation remains however poorly understood.  
Via mathematical modelling, we identify four distinct synchronization regimes depending on two quantities: the duration of the so-called licensing period during which the initiation potential in the cell remains high after the first origin has fired and the duration of the blocking period during which already initiated origins remain blocked. 
For synchronous replication initiation, the licensing period must be long enough such that all origins can be initiated, but shorter than the blocking period to prevent reinitiation of origins that have already fired. We find an analytical expression for the degree of synchrony as a function of the duration of the licensing period, which we confirm by simulations. 
Our model reveals that the delay between the firing of the first and the last origin scales with the coefficient of variation (CV) of the initiation volume. Matching these to the values measured experimentally shows that the firing rate must rise with the cell volume with an effective Hill coefficient that is at least 20; the probability that all origins fire before the blocking period is over is then at least 92\%. Our analysis thus reveals that the low CV of the initiation volume is a consequence of synchronous replication initiation. Finally, we show that the previously presented molecular model for the regulation of replication initiation in \textit{E. coli} can give rise to synchronous replication initiation for biologically realistic parameters.
\end{abstract}

\maketitle
\section{Introduction}
\begin{figure*}%[\sidecaptionrelwidth][t]
	\centering
	\includegraphics[width=0.75\linewidth]{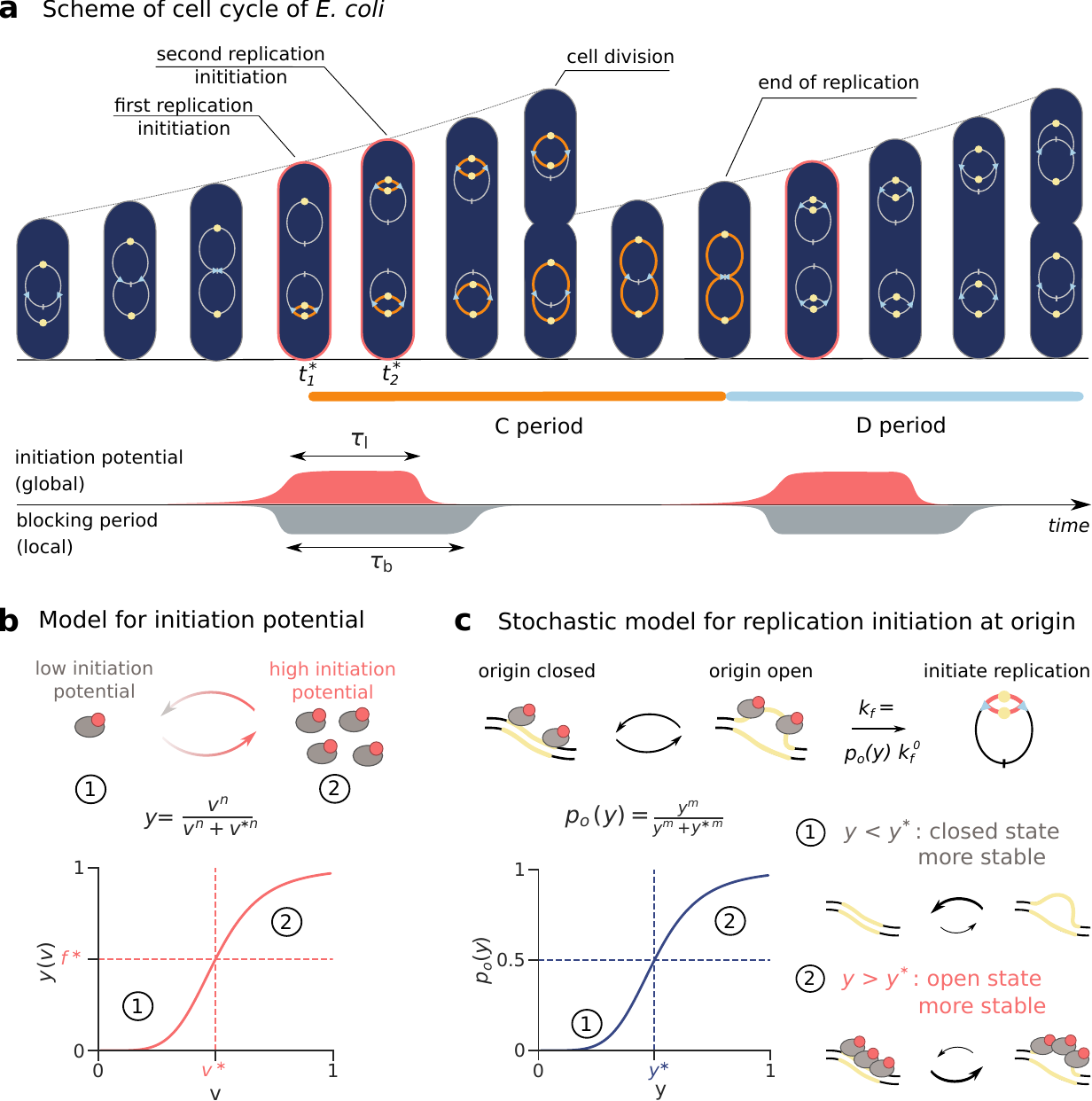}
	\caption{
		\textbf{Model of stochastic replication initiation at each origin.}
		(a) Scheme of the cell cycle of \textit{E. coli}: The volume of the cell grows exponentially with a growth rate $\lambda$. At doubling times $\tau_{\rm d}= \ln(2)/\lambda$ that are shorter than the time to replicate the entire chromosome and divide (C+D period), cells are typically born with an ongoing round of chromosomal replication. Replication is initiated stochastically at each origin (yellow circles) at times $t_1$ and $t_2$, respectively, and the  replication forks (blue triangles) advance towards the terminus (grey bar) with a constant replication speed. In \textit{E. coli}, all origins fire within a very short time interval, thus giving rise to synchronous replication initiation. To fire replication synchronously a global and a local mechanism are required: the global mechanism keeps the initiation potential high for a licensing period $\tau_{\rm l}$ (red shaded area), while the local mechanism based on SeqA prevents already initiated origins from refiring for a blocking period $\tau_{\rm b}>\tau_{\rm l}$ (grey shaded area). In our model, cell division is triggered a fixed cycling time $\tau_{\rm cc}=T_{\rm C} + T_{\rm D}$ after replication has been initiated. (b) We model the initiation potential $y$ in the cell as a function of the volume per origin $v=V / n_{\rm ori}$ via a Hill function with the Hill coefficient $n$. At the critical volume per origin $v^\ast$ the initiation potential equals the critical initiation potential $y^\ast=0.5$. (c) Stochastic model of replication initiation at the origin as a function of the initiation potential in the cell: The origin can be in an open or in a closed configuration and replication can be initiated with a constant rate $k_{\rm f}^0$ if the origin is open. The probability to be in the open state $p_{\rm o} (y)$ depends on the initiation potential in the cell and is modelled via a Hill function with the Hill coefficient $m$ and the critical active fraction of DnaA $f^\ast$. }
	\label{fig:fig_1}
\end{figure*}

Passing on the genetic information from one generation to the next with high fidelity is crucial for the survival of every organism. Many bacteria contain several 
copies of their chromosome \cite{Jain2012, Wallden2016, Si2017, Sauls2019, Santi2013, Egan2004}. 
In nutrient-rich environments, the bacterium \textit{Escherichia coli} initiates DNA replication of several copies of the same chromosome synchronously with very high precision \cite{Wallden2016, Si2017, Sauls2019, Boesen2022}. Already in the 1960s, Cooper and Helmstetter suggested that initiating new rounds of replication synchronously at several origins enables \textit{E. coli} to divide even faster than the fixed time it takes to replicate its entire chromosome \cite{Cooper1968}: Rounds of replication that started in the mother cell continue to be replicated during cell division and finish only in the following generations (Fig. \ref{fig:fig_1}a). To ensure that all daughter cells obtain a fully replicated copy of the chromosome at these high division times, replication must be initiated at all chromosomes synchronously. Later, Skarstad et al. confirmed the prediction of Cooper and Helmstetter by counting the numbers of origins in rapidly growing cultures: They found that most cells have $2^n$ ($n=1, 2, 3$) copies of their chromosome and only a small fraction of cells (2-7$\%$) contained 3, 5, 6 or 7 chromosomes \cite{Skarstad1986}. Recent single-cell measurements indeed show that \textit{E. coli} initiates replication synchronously at up to eight origins with very high precision in the fast growth regime \cite{Wallden2016, Boesen2022}. It remains however an open question how \textit{E. coli} achieves such a high degree of synchrony.

Replication initiation in \textit{E. coli} is controlled by the initiator protein DnaA \cite{Atlung1991, Hansen2018, Katayama2010, Katayama2017}. This protein can switch between two nucleotide-binding states, an inactive state in which DnaA is bound to ADP and an active one in which it is bound to ATP \cite{Wallden2015, Dewachter2018, Katayama2017, Kasho2013, Kasho2014, Kurokawa1999}. Both the inactive and active form can bind to an origin of replication, but binding of the inactive state is not sufficient: replication initiation requires the binding of ATP-DnaA \cite{Katayama2010, Nishida:2002dp, Speck2001, Keyamura2009}. The evidence is accumulating that the origin binding of DnaA and hence replication initiation is controlled via two distinct mechanisms, titration and protein activation \cite{Kurokawa1999, Katayama2017, Hansen1991}. Titration of DnaA via high-affinity DnaA binding sites on the chromosome generates a cycle in the concentration of free DnaA that is available for binding to the origin \cite{Katayama2017, Schaper1995}, while an activation switch induces a cycle in the fraction of active DnaA \cite{Kurokawa1999, Fujimitsu2009, Katayama2001}. These two cycles together conspire to generate robust oscillations in the concentration of free and active DnaA \cite{Berger2022}. This concentration of free and active DnaA forms the initiation potential of the cell, which determines the propensity of origin firing.
	
Initiation synchrony entails that all origins are initiated during each cell cycle, yet also only once per cell cycle. This is a major challenge because the cell needs to meet two potentially conflicting constraints. The requirement that all origins must fire during each cell cycle means that when the first origin fires, the initiation potential cannot go down immediately: it must continue to rise so that also the other origins can fire. On the other hand, the origin that has fired, should not fire again, even though the initiation potential is still rising. It appears that \textit{E. coli} employs two distinct mechanisms to meet these two constraints. The oscillations in the initiation potential, the concentration of free and active DnaA, constitute a global mechanism that induces not only the first origin to fire, but also prompts, and allows, the remaining origins to fire (Fig. \ref{fig:fig_1}a). To prevent the immediate reinitiation of origins that have already fired, a local mechanism is used.

The local mechanism that prevents the immediate reinitiation of newly replicated origins is based on the so-called sequestration of these origins. 
In \textit{E. coli}, after an origin has initiated replication, the protein SeqA transiently binds to this origin and thus prevents that new rounds of replication start immediately again at the same origin \cite{Riber2016, Campbell:1990it}. When either of the two proteins SeqA or Dam that are required for sequestration after replication initiation are deleted, synchrony is lost and replication is initiated throughout the entire cell cycle \cite{Riber2016, Lu:1994ee}. 
Blocking of recently initiated origins during a so-called blocking period is therefore an essential mechanism to ensure synchronous replication initiation (Fig. \ref{fig:fig_1}a). 

The combination of global oscillations in the initiation potential, which induce all origins to fire, and local origin sequestration, which prevents the newly replicated origins from reinitiation, appears to be an elegant solution to the problem of initiation synchrony. Yet, many questions remain. Newly replicated origins are only sequestered for a finite amount of time: the blocking period is about 10 minutes long \cite{Katayama2010, Lu:1994ee, Waldminghaus:2009em}. Hence, while, after the initial origin has fired, the initiation potential must first continue to rise sufficiently long in order to allow all the remaining origins to fire, it must also come down before this blocking period is over because otherwise, the newly replicated origin(s) will fire again after all. The licensing period during which origins can fire must thus be long enough for all origins to fire, yet also shorter than the blocking period (Fig. \ref{fig:fig_1}a). Given that the blocking period is only 10 minutes, this constraint is likely to pose a major challenge.

The problem of replication synchrony is compounded by the fact that the oscillations in the initiation potential are directly shaped by replication initiation itself \cite{Riber2016}. When a new origin is fired, the newly generated replisomes will stimulate the deactivation mechanism called RIDA \cite{Katayama2010, Kurokawa1999, Kato2001, Nakamura2010}. Moreover, a few minutes after an origin has initiated DNA replication, the locus \datA is duplicated, which enhances deactivation by stimulating the hydrolysis of ATP bound to DnaA \cite{Kitagawa1996, Kitagawa1998, Ogawa2002, Nozaki2009, Kasho2013}. Furthermore, the newly duplicated DNA will harbor new titration sites \cite{Schaper1995, Katayama2017}, which also tend to reduce the initiation potential by lowering the concentration of cytoplasmic DnaA. 
How these molecular mechanisms cause the initiation potential to first continue to rise during the licensing period and then fall before the blocking period is over is far from understood. 

To study how replication can be initiated synchronously at several origins, we first propose a minimal coarse-grained model in which an initiation potential rises when the cell reaches a critical volume per origin. Each origin can initiate stochastically with a firing probability that depends on the initiation potential. The model contains a licensing period during which the initiation potential rises and origins can fire, and a blocking period during which newly fired origins cannot fire again. By varying the duration of the licensing and the blocking period we reveal four regimes. Only one of these gives rise to robust synchronous replication initiation. In particular, in order to initiate synchronously, the licensing period must be long enough for all origins to fire, yet shorter than the blocking period. However, given that the measured blocking period is only 10 minutes \cite{Katayama2010, Lu:1994ee, Waldminghaus:2009em}, the licensing period must be shorter than 10 minutes. To fire all origins within this short blocking period with a success rate of 92\%, the firing rate must rise with the volume with a Hill coefficient of at least 20, such that the average time between the first and last initiation event is less than 4 minutes, as measured experimentally by Skarstad et al. \cite{Skarstad1986}.  Our modelling thus provides a rationale for the question of why DNA replication initiation in \textit{E. coli} is so tightly controlled.

We then investigate how these general synchronization requirements could be realized in the bacterium \textit{E. coli}, by replacing the coarse-grained initiation potential with our previously proposed molecular model, in which the free ATP-DnaA concentration oscillates over the course of the cell cycle \cite{Berger2022}; to this end, we have extended this model to include stochastic origin firing. We find that if replication initiation is controlled by the DnaA activation switch \cite{Kurokawa1999, Fujimitsu2009, Katayama2001, Katayama2017}, initiation synchrony is only achieved for a narrow range of parameters, which is hard to reconcile with the experimentally measured values. Adding titration \cite{Hansen1991, Schaper1995, Hansen2018} and bringing the system into a regime where the DnaA concentration in the cytoplasm is low during most of the cell cycle significantly improves the degree of synchrony by sharpening the rise of the initiation potential at a critical volume per origin. This suggests that combining a concentration cycle based on titration with a protein activation cycle is crucial for initiating replication synchronously at multiple origins in the bacterium \textit{E. coli}.

\section{The licensing period must be non-zero and shorter than the blocking period}
Our coarse-grained model to study the effect of stochastic replication initiation on the cell cycle consists of two parts: Firstly, we model the available amount of initiator proteins in the cell as an initiation potential $y$ that depends on the volume per origin $v(t)= V(t)/n_{\rm ori}(t)$ according to 
\begin{equation}
y(v) = \frac{v^{n}}{v^{n} + v^{ \ast n}}
\label{eq:initiation_potential}
\end{equation}
with the Hill coefficient $n$ and the critical volume  per origin $v^\ast$ (Fig. \ref{fig:fig_1}b). Secondly, each origin is modelled as a two-state system that can switch stochastically between an open and a closed configuration (see Appendix \ref{sec:origin_dynamics_details} for details). Exploiting that the origin is more likely to be in the open state when the initiation potential $y$ in the cell is high, we assume that the probability to be in the open state $p_{\rm o}$ increases with the activation potential $y$ as
\begin{equation}
p_{\rm o}(y) = \frac{y^{m}}{y^{m} + y^{ \ast m}}
\label{eq:opening_prob_f}
\end{equation}
with the Hill coefficient $m$ and the critical initiation potential $y^\ast$ (Fig. \ref{fig:fig_1}c). Molecularly, this non-linear opening probability $p_{\rm o}(y)$ could arise via cooperative binding of the initiator to the origin or via a Monod-Wyman-Changeux model, where the open configuration becomes more energetically favorable the more initiators bind to the origin. Assuming rapid opening and closing dynamics of the origin, the origin firing rate is given by the probability to be in the open state $p_{\rm o}$ times the maximal firing rate $k_{\rm f}^0$:
\begin{equation}
k_{\rm f} = k_{\rm f}^0 \, p_{\rm o}
\label{eq:firing_rate}
\end{equation}

\begin{figure}
	\centering
	\includegraphics[width=\linewidth]{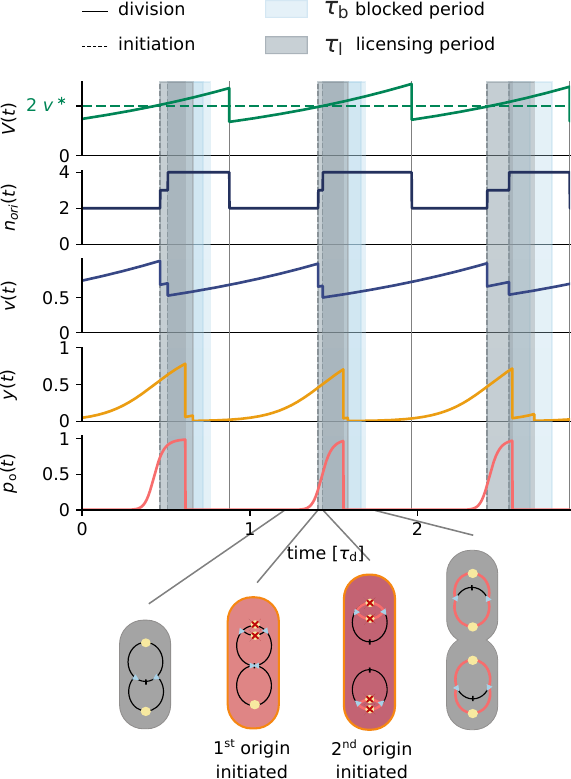}
	\caption{\textbf{Replication is initiated synchronously at several origins by introducing a blocking and a licensing period.} The volume $V(t)$, the number of origins $n_{\rm ori}(t)$, the volume per number of origins $v(t) = V(t)/ n_{\rm ori}(t)$, the initiation potential $y(t)$ and the opening probability $p_{\rm o}(t)$ as a function of time (in units of the doubling time of the cell $\tau_{\rm d}$). Every origin is initiated stochastically (dashed vertical grey lines) and during the blocking period $\tau_{\rm b}$ (light blue shaded area), the newly replicated origins cannot be re-initiated. The initiation potential $y(t)$ and the opening probability $p_{\rm o}(t)$ continue to increase during the licensing period $\tau_{\rm l}$ (grey shaded area), such that the remaining origins that have not yet initiated replication are also initiated. At the end of the licensing period, the initiation potential $y(t)$ and therefore also the opening probability $p_{\rm o}(t)$ instantaneously decreases to a lower value, making re-initiation highly unlikely. At cell division (vertical solid grey lines), the cell volume is divided by two and one of the two chromosomes is chosen at random for the next cell cycle. The cartoon below shows cells and their circular chromosome at four different moments of the cell cycle. Replication is initiated at the origin (yellow circle) and advances to the terminus (black bar) via two replication forks (light blue triangles).  Blocked origins are marked by red crosses and the shaded color of the cell indicates whether the initiation potential in the cell is low (grey color) or high (red color).
	(See Table \ref{tab:parameters} for all parameters.)
	}
	\label{fig:fig_2_1_synch}
\end{figure}

\begin{figure*}%[tbhp]
	\centering
	\includegraphics[width=0.55\linewidth]{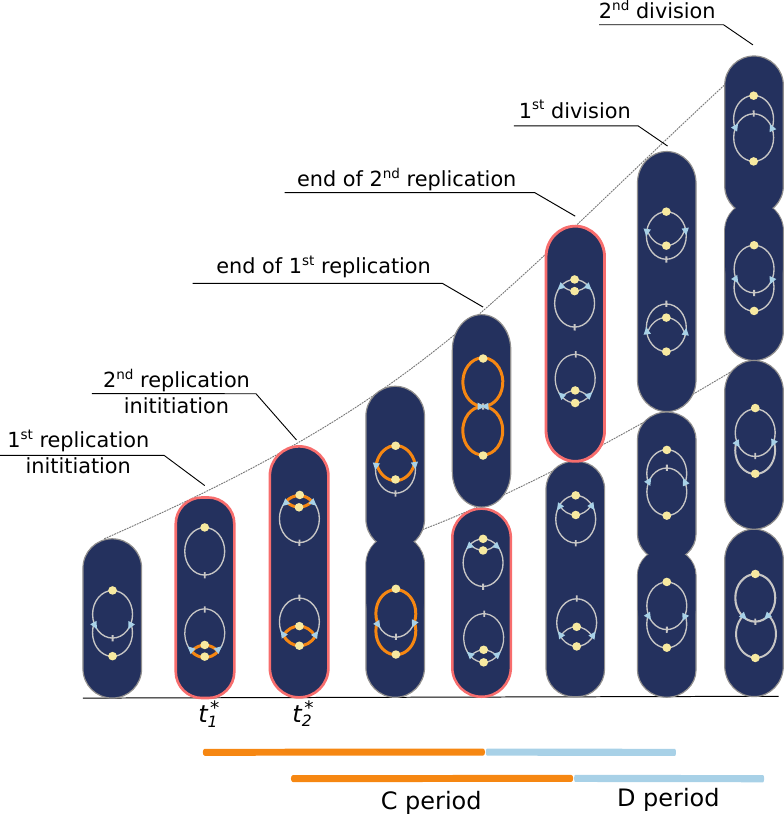}
	\caption{\textbf{Scheme of the cell cycle of \textit{E. coli} that shows how each origin firing event triggers cell division a fixed cycling time later.} At doubling times that are shorter than the time to replicate the entire chromosome and divide (C+D period), cells are typically born with an ongoing round of chromosomal replication. Here, we illustrate replication initiation in the growth-rate regime with two origins of replication at the beginning of the cell cycle. Replication is initiated stochastically at each origin (yellow circles) at times $t_1$ and $t_2$, respectively, and the replication forks (blue triangles) advance towards the terminus (grey bar) with a constant replication speed. In our model, each initiation event triggers cell division a fixed cycling time $\tau_{\rm cc}=T_{\rm C} + T_{\rm D}$ after replication has been initiated. This ensures that a cell never divides before the entire chromosome has been replicated. Note that ``1$^{\rm st}$ division'' (``2$^{\rm nd}$ division'') corresponds to the division event triggered by the ``1$^{\rm st}$ replication initiation'' (``2$^{\rm nd}$ replication initiation'') event in the mother cell. 
	}
	\label{fig:fig_A10_scheme_division}
\end{figure*}

To investigate the effect of stochastic replication initiation on the cell cycle of \textit{E. coli}, we model the volume $V(t)$ of the cell as an exponential function, $V(t)= V_{\rm b} \, e^{\lambda \, t}$, where the growth rate $\lambda = \rm{ln}(2)/\tau_{\rm d}$, with cell-doubling time $\tau_{\rm d}$, is a model parameter. We track the number of chromosomes together with their state of replication (e.g. fully replicated or replication ongoing) and whenever an origin fires a new round of replication at time $t^\ast$, a new division time a constant cycling time $\tau_{\rm cc}$ after replication initiation is set at $\tau_{\rm div} = t^\ast + \tau_{\rm cc}$. The constant cycling time $\tau_{\rm cc}$ is given by the sum of the time to replicate the entire chromosome $T_{\rm C}$ and the time from the end of replication until cell division $T_{\rm D}$ (Fig. \ref{fig:fig_1}a). Every set division time $\tau_{\rm div}$ is linked to the chromosome that just initiated replication. This choice ensures that a cell never divides before the chromosome has been replicated (Fig. \ref{fig:fig_A10_scheme_division}). When the next division time is reached, the cell volume is divided by two, and one of the two daughter chromosomes is kept at random for the next cell cycle. When the cell inherits a chromosome that is already being replicated but has not yet reached its division time, it also inherits the next division time (Fig. \ref{fig:fig_A10_scheme_division}). 

For synchronous replication initiation at several origins, the initiation potential must remain high after the first initiation event during a licensing time $\tau_{\rm l}$ and already initiated origins must be prevented from reinitiation during a blocking time $\tau_{\rm b}>\tau_{\rm l}$. To study the effect of stochastic replication initiation on the degree of synchrony, we consider the fast growth regime ($\tau_{\rm d} < \tau_{\rm cc}$), where there are typically two or more origins in the cell at the moment of replication initiation. In what follows below, we focus on the regime with two origins at the beginning of the cell cycle, yet also argue that these results generalize to regimes with more origins at the beginning of the cell cycle. At the critical volume per origin $v^\ast$, the activation potential $y(t)$ rises, and the probability to initiate replication $p_{\rm o}(t)$ increases strongly (Fig. \ref{fig:fig_2_1_synch}, lowest panel).  When the first origin fires, the number of origins in the cell increases stepwise, and the volume per origin $v(t)$ drops instantaneously (Fig. \ref{fig:fig_2_1_synch}, second and third panel). If the initiation potential $y(t)$ (and therefore also the opening probability $p_{\rm o}(y)$) followed the change in the volume per origin instantaneously, it would become very unlikely for the second origin to initiate replication as well, resulting in asynchronous replication initiation. We therefore introduce a licensing time $\tau_{\rm l}$, during which the initiation potential does not yet sense the change in the volume per origin $v(t)$ and continues to rise (Fig. \ref{fig:fig_2_1_synch}). The opening probability $p_{\rm o}(t)$ therefore rises sharply during this licensing time and the second origin also initiates replication stochastically. In order to prevent that already initiated origins fire again, we additionally introduce a blocking period $\tau_{\rm b}$, during which replication cannot be initiated again at the same origin (Fig. \ref{fig:fig_2_1_synch}, red crossed origins in the cartoon). At the end of the licensing time, the activation potential is updated according to the current number of origins in the cell, and it thus drops instantaneously (Fig. \ref{fig:fig_2_1_synch}, fourth panel).  For a sufficiently long licensing time and a block period that is longer than the licensing time, $\tau_{\rm b}>\tau_{\rm l}$, we indeed obtain stable cell cycles with synchronous replication initiation events (Fig. \ref{fig:fig_2_1_synch}). 

\begin{figure*}
	\centering
	\includegraphics[width=0.9\linewidth]{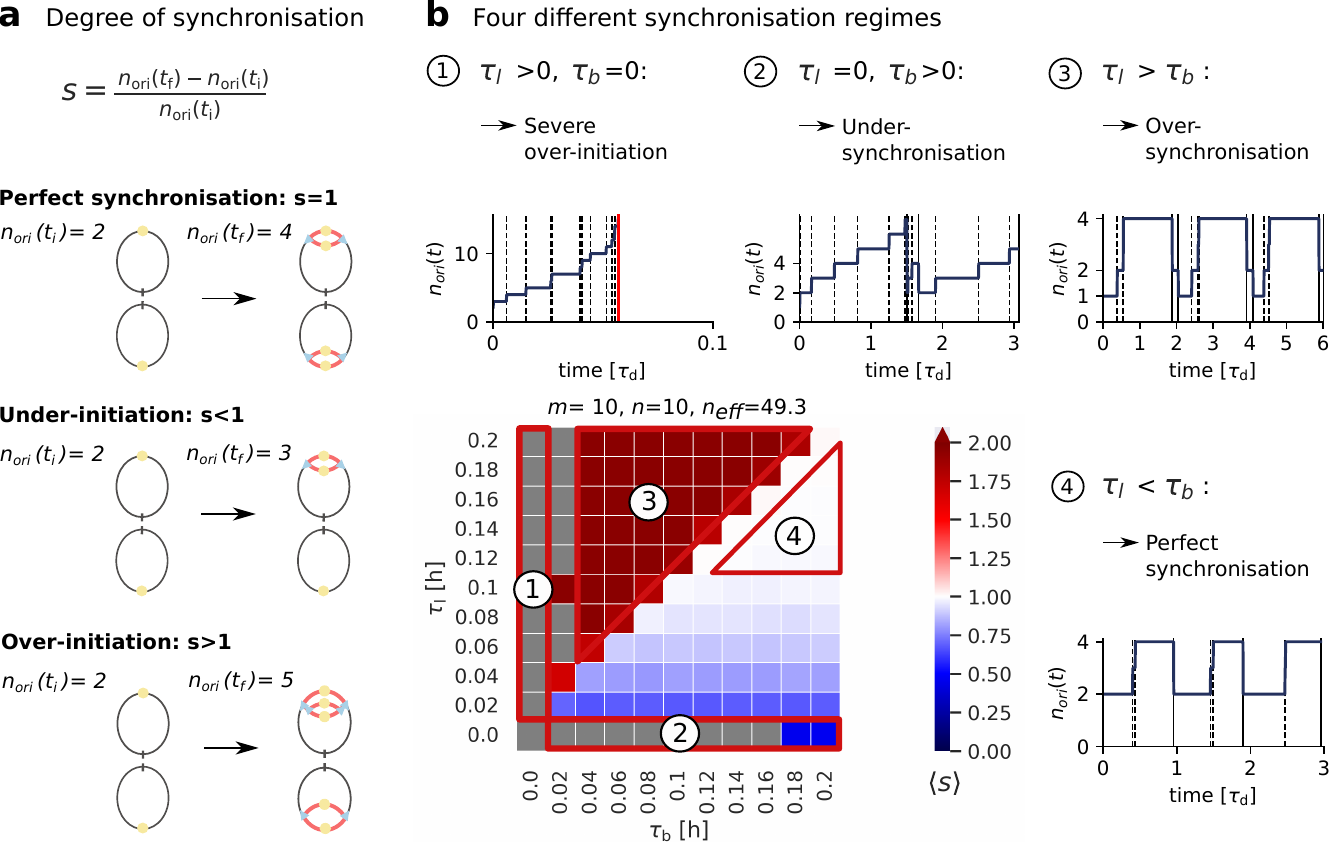}
	\caption{\textbf{Replication is only initiated synchronously when the licensing period is sufficiently long, yet shorter than the blocking period.} (a) The degree of synchrony $s$ of an initiation cascade is given by the number of origins at the end of the initiation cascade $n_{\rm ori}(t_{\rm f})$ minus the number of origins at the beginning of the initiation cascade $n_{\rm ori}(t_{\rm i})$ relative to the number of origins at the beginning of the initiation cascade $n_{\rm ori}(t_{\rm i})$. An initiation cascade begins with the moment where the first origin fires and ends after the licensing time $\tau_{\rm l}$. Replication is initiated synchronously when all origins that were present at the beginning of the cascade have fired exactly once during the cascade ($s=1$) and replication was under- or overinitiated when less or more origins have initiated, respectively. (b) The average degree of synchrony $\langle s \rangle$ as a function of the licensing period $\tau_{\rm l}$ and the blocking period $\tau_{\rm b}$. The effective Hill coefficient $n_{\rm eff}$ is obtained by fitting the opening probability $p_{\rm o}(f(v))$  to a Hill function $p_{\rm o}(v)$ (Appendix \ref{sec:methods_deriv_eff_hill_coeff}). For each parameter set, the average degree of synchrony was obtained from $N=5000$ consecutive cell cycles. We show example time traces of the number of origins as a function of time (in units of the doubling time of the cell $\tau_{\rm d}$) for four different synchronization regimes as indicated in the heatmap. When no cell cycle could be obtained, the field in the heatmap is marked in grey.
		(See Table \ref{tab:parameters} for all parameters.)
	}
	\label{fig:fig_2_2_synch}
\end{figure*}

To quantify the degree of synchrony of replication initiation for a given parameter set, we define the degree of synchrony $s$ as the change of the number of origins $\Delta n_{\rm ori}$ from the beginning of the initiation period $t_{\rm i}$ to the end of the initiation period $t_{\rm f}$, relative to the number of origins $n_{\rm ori}(t_{\rm i})$ at the beginning of the initiation period (Fig. \ref{fig:fig_2_2_synch}a):
\begin{equation}
s = \frac{\Delta n_{\rm ori}}{n_{\rm ori}(t_{\rm i})} = \frac{n_{\rm ori}(t_{\rm f}) - n_{\rm ori}(t_{\rm i})}{n_{\rm ori}(t_{\rm i})}
\label{eq:degree_of_synch}
\end{equation}
The beginning of the initiation period $t_{\rm i}$ is given by the time at which the first origin fires and the initiation period ends at $t_{\rm f} = t_{\rm i} + \tau_{\rm l}$, when the licensing period of the first origin that has fired is over.
When the degree of synchrony $s$ is one, replication is initiated synchronously, as all origins that were present at the beginning of the initiation period have fired (Fig. \ref{fig:fig_1}a). For $s<1$ or $s>1$, replication is under- or over-initiated, respectively (Fig. \ref{fig:fig_2_2_synch}a). We measure the degree of synchrony $s$ for many cell cycles to obtain the average degree of synchrony $\langle s \rangle$ for any parameter set.

By varying the duration of the licensing and the blocking period, we find four different synchronization regimes (Fig. \ref{fig:fig_2_2_synch}). All simulations start with a single, fully replicated chromosome and if replication is initiated in perfect synchrony, the system settles to a state, where the number of origins oscillates between two and four. However only in regime four is replication initiated synchronously (Fig. \ref{fig:fig_2_2_synch}b, regime 4).
When the blocking period $\tau_{\rm b}$ is zero but the licensing period $\tau_{\rm l}$ is larger than zero, replication is severely over-initiated, such that no stable cell cycles can be obtained (Fig. \ref{fig:fig_2_2_synch}b, regime 1, grey fields). If on the other hand the licensing period $\tau_{\rm l}$ is zero or very short and the blocking period $\tau_{\rm b}$ is larger than $\tau_{\rm l}$, we obtain a highly under-synchronized cell-cycle: After each initiation event, the initiation potential drops rapidly, preventing further initiation events. This results in periodic, individual initiation events throughout the entire cell cycle (Fig. \ref{fig:fig_2_2_synch}b, regime 2). When both the licensing and the blocking period are non-zero and the licensing period is longer than the blocking period, $\tau_{\rm l}> \tau_{\rm b}$, origins that have already fired can fire again after the end of the licensing period. This results in initiation events where all origins fire synchronously, but with too many origin firing events. As can be seen in Fig. \ref{fig:fig_2_2_synch}b, in this third regime the number of origins goes from one to four during one initiation duration instead of oscillating between two and four. We therefore call this regime ``over-synchronized''.  Replication is only initiated synchronously once per cell cycle when the licensing period $\tau_{\rm l}$ is sufficiently large and the blocking period is even larger $\tau_{\rm b} >\tau_{\rm l}$ (Fig. \ref{fig:fig_2_2_synch}b, regime 4). 
\section{A steep rise in the origin opening probability is essential}
A key question is what controls the size of the synchronization regime 2. While the transition from the over-initiation (Fig. \ref{fig:fig_2_2_synch}b, regime 3) to the perfect synchronization regime (Fig. \ref{fig:fig_2_2_synch}b, regime 4) is sharp and clearly defined by the requirement that $\tau_{\rm b} >\tau_{\rm l}$, the transition from the under-synchronization (Fig. \ref{fig:fig_2_2_synch}b, regime 2) to the perfect synchronization regime (Fig. \ref{fig:fig_2_2_synch}b, regime 4) is smooth and there is no clear separation between these two regimes.
In the following, initially still focusing on the regime with two origins at the beginning of the cell cycle, we show that the average degree of synchrony $\langle s \rangle$ at different delay periods $\tau_{\rm l}$ and Hill coefficients $n$ and $m$ can be derived from the probability that two independent origins fire within a time interval $\Delta t< \tau_{\rm l}$. 

To calculate the probability that two independent firing events happen within a time interval $\Delta t$, we first derive an analytical expression for the instantaneous firing probability $k_{\rm f}(t)=k_{\rm f}^{0} \, p_{\rm o}(t)$.
In our model, the opening probability $p_{\rm o}(y)$ depends indirectly on the time-dependent volume per origin $v(t)$ via the activation potential $y(v)$, see equations \ref{eq:initiation_potential} and \ref{eq:opening_prob_f}. The opening probability as a function of the volume per origin $v$ can however be approximated by a Hill function (see Appendix \ref{sec:methods_deriv_eff_hill_coeff} for derivation)
\begin{equation}
p_{\rm o}(v) \approx \frac{v^{n_{\rm eff}}}{v^{n_{\rm eff}} + v^{\ast n_{\rm eff}}}
\label{eq:opening_prob_of_v_approx_final}
\end{equation}
with the effective Hill coefficient
\begin{equation}
n_{\rm eff}=\frac{n \, m}{2}.
\label{eq:n_eff_approx_definition}
\end{equation}   
This is a good approximation for the opening probability $p_{\rm o}(y(v))$, when both the Hill coefficient of the activation potential and that of the opening probability, $n$ and $m$, respectively, are relatively high (see Eqs. \ref{eq:initiation_potential} and \ref{eq:opening_prob_f} and Fig. \ref{fig:fig_A1}c and d). The firing rate is then given by equation \ref{eq:firing_rate} with the approximate opening probability $p_{\rm o}(v)$ from equation \ref{eq:opening_prob_of_v_approx_final}. In the following, we use the procedure proposed in Ref. \cite{Wallden2016} where the maximal firing rate $k_{\rm f}^0$ is chosen such that the average initiation volume $\langle v^\ast \rangle$ equals the theoretical initiation volume $v^\ast$ in equation \ref{eq:opening_prob_of_v_approx_final} (See Appendix \ref{sec:parameter_choice_max_opening_rate}).
Using this analytical approximation for the opening probability in the regime of sufficiently high Hill coefficients $n$ and $m$, we can now calculate the probability that two independent initiation events at times $t_{1}$ and $t_{2} > t_{1}$ happen within a time interval $\Delta t = t_2-t_1 \le \tau_{\rm l}$ (Appendix \ref{sec:derivation_degree_synch_th}). In order to compare this probability $\langle P(\Delta t < \tau_{\rm l})\rangle$ to the degree of synchrony obtained from the simulations in the growth regime where two origins are present at the beginning of an initiation event, we re-scale the probability to range from $s_{\rm min}= 0.5$ to $s_{\rm max}= 1$:
\begin{equation}
s_{\rm th} = 0.5 + \langle P(\Delta t < \tau_{\rm l})\rangle \times 0.5
\label{eq:degree_synchr_th}
\end{equation}

\begin{figure*}%[tbhp]
	\centering
	\includegraphics[width=0.75\linewidth]{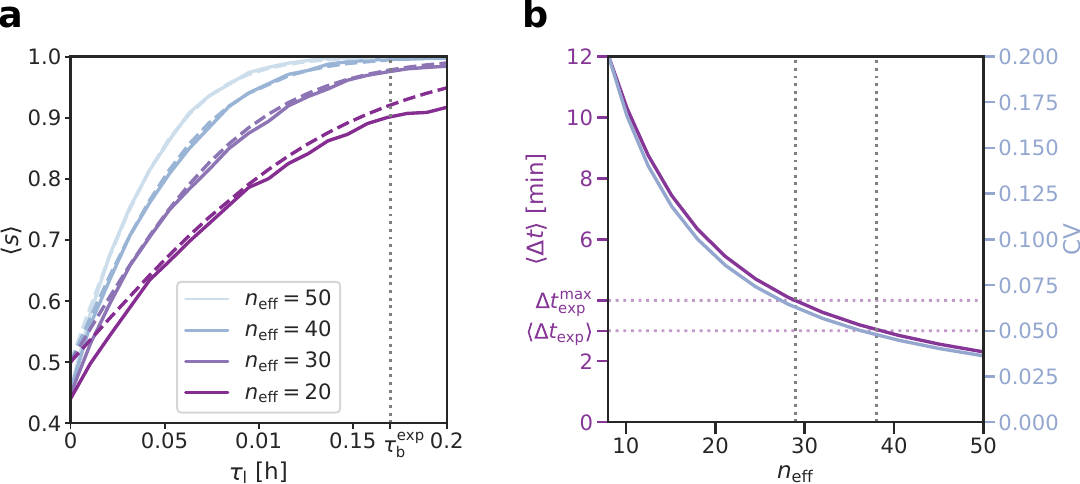}
	\caption{\textbf{The experimentally observed high precision of replication initiation is required to ensure synchronous replication initiation at multiple origins.}  
		(a) The average degree of synchrony $\langle s \rangle$ as a function of the licensing time $\tau_{\rm l}$ for varying effective Hill coefficients $n_{\rm eff}$ (with $n=m=\sqrt{2 \, n_{\rm eff}}$) from the simulations (solid lines) agrees well with the theoretical prediction derived in the Appendix \ref{sec:derivation_degree_synch_th} (dashed lines). 
		The small difference between the simulations and theory at very low delay periods arises from the fact that while in the theory for two synchronous firing events, the minimal degree of synchrony is $s_{\rm min}=0.5$, in the simulations there can be more origins at the beginning of an initiation cascade, leading to a lower degree of synchrony $s_{\rm min}<0.5$.
		In these simulations, the blocking period $\tau_{\rm b}$ is set larger than all tested licensing $\tau_{\rm l}$ periods ($\tau_{\rm b}=0.25 \, {\rm h}$), such that over-initiation events cannot occur. The maximal firing rate $k_{\rm f}^0$ is set such that the average initiation volume $\langle v^\ast \rangle$ equals the theoretical initiation volume $v^\ast$ in equation \ref{eq:opening_prob_of_v_approx_final} as explained in Appendix \ref{sec:parameter_choice_max_opening_rate}. The experimentally measured blocking period $\tau_{\rm b}^{\rm exp}$ is marked as a grey vertical dotted line.
		For each parameter set, the average degree of synchrony was obtained from $N=5000$ consecutive cell cycles. 
		(b) The theoretical average time interval between two consecutive firing events $\langle \Delta t \rangle$ (pink line and axes) and the coefficient of variation of the initiation volume (CV, blue line and axes) as a function of the effective Hill coefficient $n_{\rm eff}$ (see Appendix \ref{sec:deriv_average_initiation_interval_cv} for derivation). Skarstad et al. \cite{Skarstad1986} found experimentally that the average time interval to fire all origins in the B/r A \textit{E. coli} strain is about $\langle \Delta t_{\rm exp} \rangle =$3~min with an upper estimate of $\Delta t_{\rm exp}^{\rm max} =$4~min (horizontal dotted pink lines). The effective Hill coefficient lies therefore in the range $n_{\rm eff}=29-38$ (vertical dotted grey lines), corresponding to a coefficient of variation of CV=0.05-0.07. This agrees well with the experimentally measured precision of the initiation volume of CV$\le$0.1 \cite{Wallden2016, Si2019, Boesen2022}. Interestingly, the average degree of synchrony at $n_{\rm eff}=30$ and $n_{\rm eff}=40$, respectively, is given by $\langle s \rangle (n_{\rm eff}=30)= 0.975$ and $\langle s \rangle (n_{\rm eff}=40)= 0.996$, corresponding to the probabilities to fire all origins synchronously of $\langle P(\Delta t < \tau_{\rm l})\rangle (n_{\rm eff}=30) \equiv P_{\rm s}(n_{\rm eff}=30)=95.5 \%$ and $P_{\rm s}(n_{\rm eff}=40)=98.9 \%$. This prediction of the degree of synchrony agrees well with the qualitative experimental observation that in \textit{E. coli} DNA replication is typically initiated synchronously at multiple origins. Our model therefore provides a rationale for the experimentally observed high precision of replication initiation.
		(See Table \ref{tab:parameters} for all parameters.)}
	\label{fig:fig_3_synch}
\end{figure*}

The average degree of synchrony $\langle s \rangle$ as a function of the licensing period $\tau_{\rm l}$ for different Hill coefficients $n$ and $m$ is indeed very well approximated by $s_{\rm th}$ (Fig. \ref{fig:fig_3_synch}a). The transition from the under-synchronized to perfect synchronization regime in Figure \ref{fig:fig_2_2_synch}b is therefore given by the probability that two independent origin firing events happen within a short time window given by the licensing time $\tau_{\rm l}$. The higher the effective Hill coefficient $n_{\rm eff}$, the higher the degree of synchrony for a given delay period $\tau_{\rm l}$ (Fig. \ref{fig:fig_3_synch}a). The degree of synchrony $\langle s \rangle$ increases with the effective Hill coefficient because that raises the firing rate more steeply, making it more likely that the two origins fire within the licensing time $\tau_{\rm l}$.

While the degree of synchrony $\langle s \rangle$ increases with $\tau_{\rm l}$, $\tau_{\rm l}$ cannot be longer than the blocking period $\tau_{\rm b}$, because otherwise, origins that have already fired will fire again. The blocking period thus bounds $\tau_{\rm l}$. In the bacterium \textit{E. coli}, the origin of the blocking period $\tau_{\rm b}$ is well understood: The protein SeqA can bind to specific sites on the newly replicated origins which overlap with the binding sites for the initiation protein and thus prevent it from reinitiating. After about ten minutes, the protein SeqA unbinds and new rounds of replication can start again \cite{Katayama2010, Lu:1994ee, Waldminghaus:2009em}.
As the licensing time must be shorter than the blocking period to prevent over-initiation (Fig. \ref{fig:fig_2_2_synch}b, regime 3), the licensing time in \textit{E. coli} must be less than $\tau_{\rm b}^{\rm exp}=$10 min. Fig. \ref{fig:fig_3_synch}a shows this puts a major constraint on the Hill coefficient: to get a degree of synchrony $\langle s \rangle$ that is above 95\%, the effective Hill coefficient must be at least $n_{\rm eff}=30$.

\begin{figure*}%[tbhp]
\centering
\includegraphics[width=0.4\linewidth]{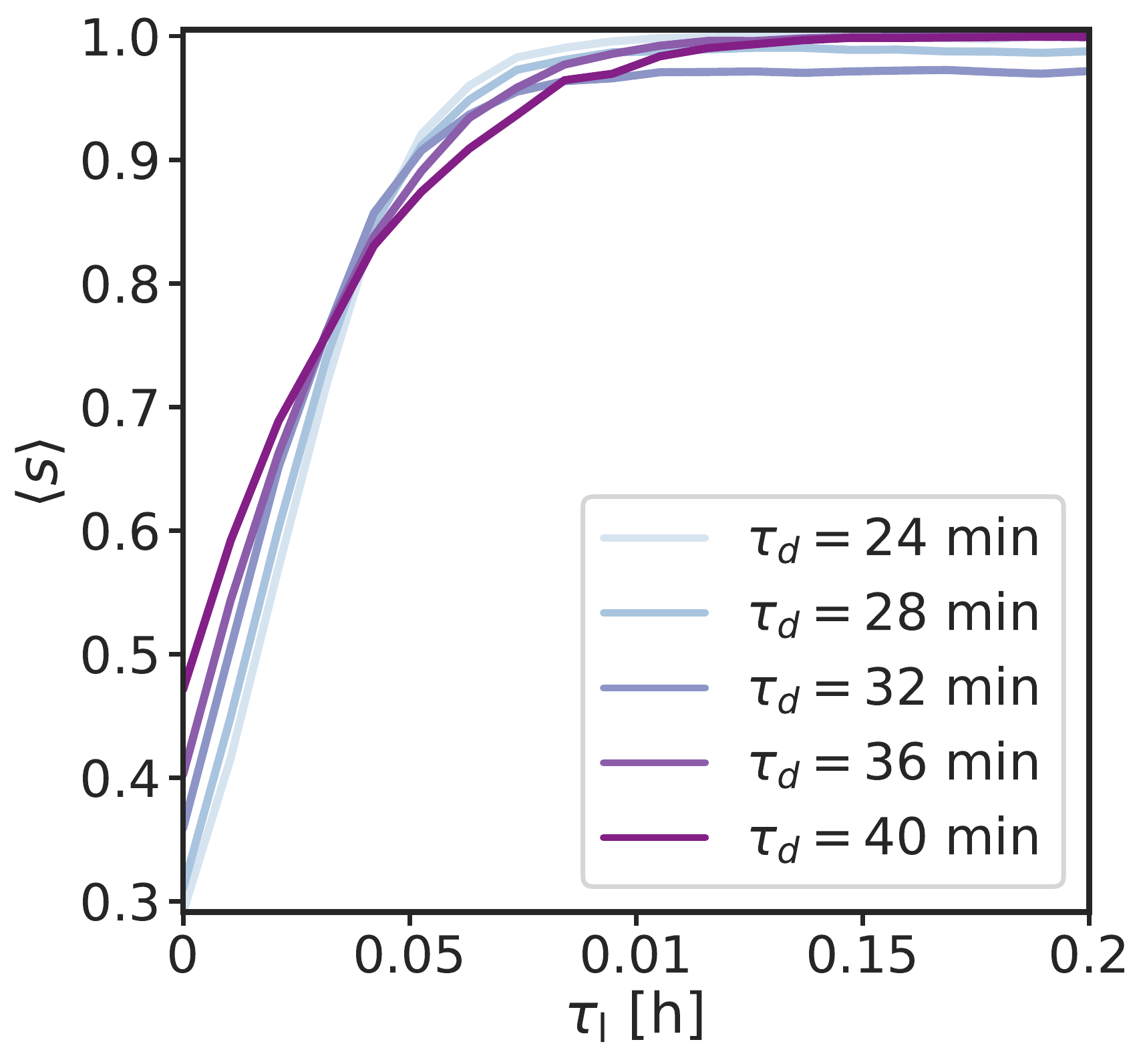}
\caption{\textbf{The average degree of synchrony does not depend strongly on the growth rate of the cell.} The average degree of synchrony $\langle s \rangle$ as a function of the duration of the licensing period $\tau_{\rm l}$ for different cell-doubling times $\tau_{\rm d}$. The average degree of synchrony $\langle s \rangle$ is to a good approximation independent of the doubling time $\tau_{\rm d}$. The small decrease in the degree of synchrony at high licensing times $\tau_{\rm l}$ at a doubling time of $\tau_{\rm d}=32$~min is because at this doubling time replication initiation happens almost at the same time as cell division. When the cell divides during an initiation cascade, the number of origins decreases and the counted total number of origins at the end of the cascade is smaller than the total number of origins in both daughter cells. This error in counting the total change in the number of origins $\Delta n_{\rm ori}$ in equation \ref{eq:degree_of_synch} effectively reduces the degree of synchrony, especially for long initiation durations (high licensing times $\tau_{\rm l}$). The decrease in the minimal average degree of synchrony $\langle s \rangle$ (at $\tau_{\rm l}=0$) with increasing growth rate comes from the overall higher number of origins at higher growth rates and thus a higher denominator in equation \ref{eq:degree_of_synch}. The effective Hill coefficient is set to $n_{\rm eff}=50$. (See Table \ref{tab:parameters} for all parameters.)}
\label{fig:fig_A3}
\end{figure*}

The question remains what the effective Hill coefficient $n_{\rm eff}$ is that is consistent with experiments. Interestingly, Skarstad et al. have measured the time between the first and last firing event, which we can compare against our theoretical prediction \cite{Skarstad1986}. However, to do so, we must first examine the dependence on the growth rate, because Skarstad performed their measurements at a higher growth rate. Fig. \ref{fig:fig_A3} shows that while the average degree of synchrony $\langle s \rangle$ as a function of the licensing time $\tau_{\rm l}$ varies strongly with the effective Hill coefficient $n_{\rm eff}$, it is fairly independent of the doubling time of the cell $\tau_{\rm d}$.

Given that $\langle s \rangle$ as a function of $\tau_{\rm l}$ is fairly independent of the growth rate, we now examine the data of Skarstad et al. \cite{Skarstad1986}. To obtain an experimental estimate for the effective Hill coefficient and thus for the average degree of synchrony, we calculate the average time interval between the first and last initiation event $\langle \Delta t \rangle$ (see Appendix \ref{sec:deriv_average_initiation_interval_cv}) and compare it to the experiments. Skarstad et al. find that this time is on average $\langle \Delta t_{\rm exp} \rangle \approx 3$~min with an upper limit of $\Delta t_{\rm exp}^{\rm max} \approx 4$~min \cite{Skarstad1986}. Our theory predicts that to fire two initiation events within an average time interval of $\langle \Delta t \rangle = 3-4$~min, the effective Hill coefficient must be in the range $n_{\rm eff}=29-38$ (Fig. \ref{fig:fig_3_synch}b, vertical grey dotted lines). Interestingly, the dependence of $\langle \Delta t \rangle$ on $n_{\rm eff}$ closely tracks that of the coefficient of variation (CV) of the initiation volume (Fig. \ref{fig:fig_3_synch}b). The $\langle \Delta t \rangle = 3-4$~min measured by Skardstadt et al. corresponds to a coefficient of variation (CV) of the initiation volume CV$\approx 0.05-0.06$. This agrees fairly well with the experimental finding that the initiation volume is one of the most tightly controlled cell cycle parameters with CV=0.08-0.1 \cite{Wallden2016, Si2019}. A CV of 0.1 as measured by Ref. \cite{Wallden2016} corresponds to $n_{\rm eff}\approx20$ (Fig. \ref{fig:fig_3_synch}b, see also calculation in Ref. \cite{Wallden2016}) and would thus only result in a relatively low degree of synchrony of less than $\langle s \rangle = 0.92$ corresponding to a probability of initiating synchronously of $\langle P(\Delta t < \tau_{\rm l})\rangle \equiv P_{\rm s}=84\%$ (see equation \ref{eq:degree_synchr_th}). Recent experiments show however that the contribution from the intrinsic noise in replication initiation to the CV is only about CV$_{\rm int}$=0.04 - 0.05 \cite{Boesen2022}, in even better agreement with the Skarstadt data (Fig. \ref{fig:fig_3_synch}b). Our model, which only concerns the effect of intrinsic noise, then predicts that for this low CV$_{\rm int}$ the effective Hill coefficient $n_{\rm eff}$ must be at least 40 (Fig. \ref{fig:fig_3_synch}b), which then means that at least $P_{\rm s}=$98\% of the initiation events happen synchronously within a period of 10 min corresponding to $\langle s \rangle = 0.99$ (Fig. \ref{fig:fig_3_synch}a).

\begin{figure*}%[tbhp]
	\centering
	\includegraphics[width=0.85\linewidth]{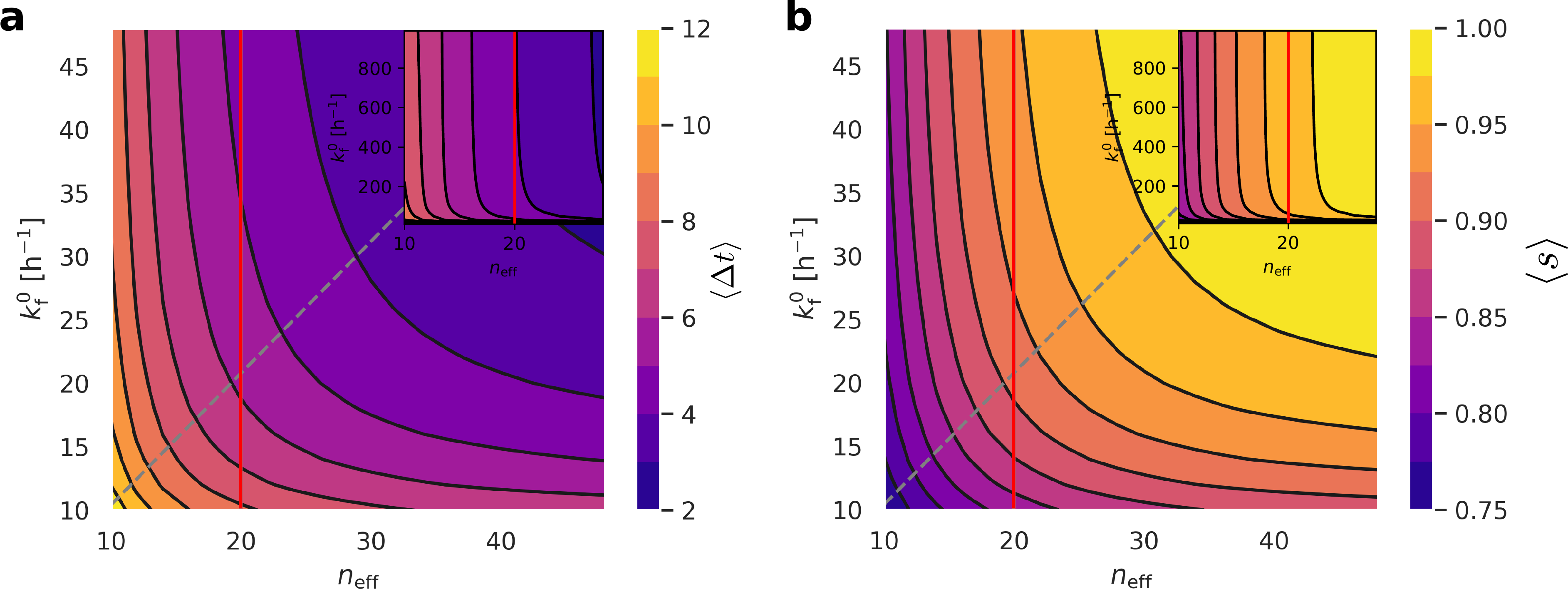}
	\caption{\textbf{In the coarse-grained model, the average time interval between the first and the last firing event $\langle \Delta t \rangle$ and the average degree of synchrony $\langle s \rangle$ depend both on the effective Hill coefficient $n_{\rm eff}$ and on the maximal firing rate $k_{\rm f}^0$.}  These contour plots show the average time interval between the first and last origin firing event $\langle \Delta t \rangle$ (a) and the theoretical degree of synchrony $s_{\rm th}$ (according to equation \ref{eq:degree_synchr_th}) (b) as a function of the effective Hill coefficient $n_{\rm eff}$ and the maximal firing rate $k_{\rm f}^0$ for the licensing period $\tau_{\rm l}=9.6$~min$<\tau_{\rm b}$ at the experimentally observed blocking period of $\tau_{\rm b}=10$~min. The dashed grey lines are given by equation \ref{eq:parameter_choice_firing_rate} and correspond to the parameter choice in the main text where the average initiation volume $\langle v^\ast \rangle=v^\ast$ (see Appendix \ref{sec:parameter_choice_max_opening_rate}). For a given Hill coefficient $n_{\rm eff}$ both $\langle \Delta t \rangle$ and $\langle s \rangle$ first increase as a function of the maximal firing rate $k_{\rm f}^0$ and then converge to a constant value for higher maximal firing rates. Therefore, at a higher maximal origin firing rate than in Fig. \ref{fig:fig_3_synch}a  the effective Hill coefficient can be lower to achieve the same degree of synchrony. However, panel a shows (and the inset more clearly) that there is a minimal $n_{\rm eff}$ necessary to reach the experimentally reported maximal bound on $\langle \Delta t^{\rm max}_{\rm exp} \rangle$, corresponding to the limit $k_{\rm f}^0 \rightarrow \infty$. The red vertical lines show that to initiate replication within at least $\langle \Delta t^{\rm max}_{\rm exp} \rangle=4$~min, the minimal Hill coefficient required in the regime of firing rates ($k_{\rm f}^0 \rightarrow \infty$) is $n_{\rm eff}=20$. This corresponds to a high degree of synchrony of $s_{\rm th} \approx 0.96$ (corresponding to $P_{\rm s} \approx 92\%$) (see panel b). Therefore, the finding of the main text that a high Hill coefficient is required for a high degree of synchrony remains valid. (See Table \ref{tab:parameters} for all parameters.)}
	\label{fig:fig_A11_isolines}
\end{figure*}

Before we conclude, we must discuss one key parameter, which is the maximal firing rate $k_{\rm f}^0$. In our theoretical model, we covaried $k_{\rm f}^0$ with $n_{\rm eff}$ to keep the average initiation volume per origin $\langle v^\ast \rangle$ constant and equal to $v^\ast$ of Eq. \ref{eq:opening_prob_of_v_approx_final}, following the procedure of Wallden et al. \cite{Wallden2016}. 
Fig. \ref{fig:fig_A11_isolines}a shows $\langle \Delta t \rangle$ in our theoretical model as a function of $k_{\rm f}^0$ and $n_{\rm eff}$ separately (thus without enforcing the constraint $\langle v^\ast \rangle=v^\ast$). While $\langle \Delta t\rangle$ increases with both $k_{\rm f}^0$ and $n_{\rm eff}$, there is a minimal $n_{\rm eff}$ that is necessary to reach a given $\langle \Delta t\rangle$, corresponding to the limit $k_{\rm f}^0 \rightarrow \infty$ (see inset Fig. \ref{fig:fig_A11_isolines}a). 
The Hill coefficient necessary to reach the $\langle \Delta t\rangle$ that matches the value measured by Skarstadt et al. is lower than that in the above procedure in which $k_{\rm f}^0$ and $n_{\rm eff}$ are covaried (corresponding to the diagonal in Fig. \ref{fig:fig_A11_isolines}a), but it is still very high, around $n_{\rm eff} \approx 20$ (Fig. \ref{fig:fig_A11_isolines}a). 
Fig. \ref{fig:fig_A11_isolines}b shows that in this limit, $k_{\rm f}^0 \rightarrow \infty$ and $n_{\rm eff}=20$, the degree of synchrony is very high, with $P_{\rm s} = 92\%$. Our model of stochastic replication initiation thus provides a rationale for the experimentally observed high precision of DNA replication initiation in \textit{E. coli}: Given the constraint set by the duration of the blocking period, the system requires a very high Hill coefficient in order to initiate replication synchronously. Since increasing the Hill coefficient beyond this already large value becomes progressively harder, it seems that the system operates close to what is theoretically possible given the duration of the blocking period.

\section{Initiation synchrony in molecular activation switch model for \textit{E. coli}}
\begin{figure*}%[tbhp]
	\centering
	\includegraphics[width=0.7\linewidth]{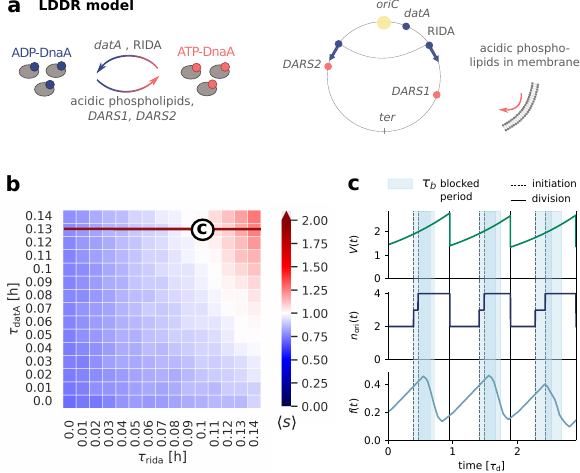}
	\caption{\textbf{The LDDR model can ensure a high degree of synchronous replication initiation for a narrow range of parameters.} (a) In the Lipid-\textit{DatA}-\textit{DARS1/2}-RIDA (LDDR) model, replication forks overlap and RIDA is the main deactivator in combination with the activators \textit{DARS1} and \textit{DARS2}. (b) The average degree of synchrony $\langle s \rangle$ as a function of the replication time of the site \datA $\tau_{\rm datA}$ and the onset time of RIDA $\tau_{\rm rida}$. The sites \textit{DARS1} and \textit{DARS2} are replicated at the experimentally measured times $\tau_{\rm d1}=0.25 \, {\rm h}=15$~min and $\tau_{\rm d1}=0.4 \, {\rm h}=24$~min, respectively. Replication is only initiated synchronously for a small range of parameters: When the site \datA is replicated after the experimentally measured time of $\tau_{\rm datA}=0.13 \, {\rm h}\approx 8$~min (red horizontal line), replication in the LDDR model is only initiated synchronously if RIDA starts only about 6 minutes after the origin is initiated. It is however not clear what could cause a delay of 6 minutes in the onset of RIDA. 
		(c) The volume $V(t)$, the number of origins $n_{\rm ori}(t)$ and the ATP-DnaA fraction $f(t)$ as a function of time (in units of the doubling time of the cell $\tau_{\rm d}$) for the parameter combination marked in b. The large amplitude oscillations in the active fraction in combination with a long delay in the onset of deactivation via RIDA and \datA can give rise to a high degree of synchrony for a small range of parameters.
		For each parameter set in b, the average degree of synchrony was obtained from $N=5000$ consecutive cell cycles. (See Table \ref{tab:parameters} for all parameters.)}
	\label{fig:fig_6}
\end{figure*}

Our coarse-grained model of replication initiation revealed general requirements for initiating replication synchronously at several origins.
It remains however an open question how these requirements are implemented on a molecular level in different organisms.
In \textit{E. coli}, both a protein activation cycle and a concentration cycle are required for robust replication initiation at all growth rates \cite{Berger2022}. 
In the following, we first address the question whether a protein activation cycle alone, i.e. without the help of a concentration cycle, can yield synchronous replication. To this end, we will study the so-called LDDR model, which we have developed previously \cite{Berger2022} (Fig. \ref{fig:fig_6}a).
This model contains activation of DnaA via the lipids and the chromosomal sites DARS1/2 and deactivation of DnaA via the chromosomal site datA and the replication-associated mechanism of Regulatory Inactivation of DnaA (RIDA) \cite{Berger2022}. We show that this cycle alone can induce synchronous replication initiation, but only over a very limited parameter regime. In a second step, we show that adding a concentration cycle via titration sites can significantly enhance the degree of synchrony.

To test the effect of stochastic origin firing in the LDDR model, we replace the abstract initiation potential $y(v)$ we used in the coarse-grained model with the LDDR model for the ATP-DnaA fraction $f(t)$ in the cell. The opening probability $p_{\rm o}(f)$ is again modelled as a simple Hill function according to equation \ref{eq:opening_prob_f}. Motivated by the experimental observation that there are 10 sites for DnaA binding to the origin \cite{Katayama2017}, the Hill coefficient was chosen to be $m = 10$; in addition, the critical fraction in Eq. \ref{eq:opening_prob_f} was chosen to be $f^\ast=0.5$. Moreover, the maximal firing rate $k_{\rm f}^0$ was set to a large value, i.e. $k_{\rm f}^0=1000$~h$^{-1}$, such that the system is in the regime where the degree of synchrony is not limited by $k_{\rm f}^0$, but only limited by the dynamics of the activation cycle $f(t)$. 
Like in the coarse-grained model, already initiated origins are blocked transiently during a blocking period of $\tau_{\rm b}=$10 min. Contrary to the coarse-grained model, where after the end of the licensing period the initiation potential drops instantaneously to a very low value, in the LDDR model, the active fraction $f$ follows from the temporal dynamics of the antagonistic interplay between DnaA activation and deactivation. In the LDDR model, the licensing period is thus not imposed, as in the coarse-grained model above, but is implicit in the dynamics of the LDDR model. Yet, to quantify the degree of synchrony, we need to define an effective initiation period $\tau_{\rm i}$, akin to the licensing period $\tau_{\rm l}$ in the coarse-grained model (see Eqs. \ref{eq:degree_of_synch} and \ref{eq:degree_synchr_th}). We define $\tau_{\rm i}$ to be a fraction of the cell cycle time $\tau_{\rm d}$: $\tau_{\rm i}=\alpha \, \tau_{\rm d}$. While $\alpha$ cannot be defined uniquely, we show in the Appendix \ref{sec:definition_cascade_duration} that the degree of synchrony is fairly robust to the precise choice of $\alpha$. In the following, we therefore choose $\alpha = 0.4$, such that $\tau_{\rm i}=0.4 \, \tau_{\rm d}$.

The LDDR model can indeed give rise to synchronous replication initiation at multiple origins, but only for a small range of parameters: When the (de)activators \textit{DARS1}, \textit{DARS2} and \textit{datA} are located at the experimentally measured positions on the chromosome, replication is initiated asynchronously when RIDA starts directly after an origin has fired (Fig. \ref{fig:fig_6}b at $\tau_{\rm rida}=0$~h). As RIDA is a strong deactivator, it causes the active fraction to drop rapidly after the first origin has been initiated and thus prevents other origins from firing as well. By varying both the position of \datA on the chromosome and the time at which RIDA starts after an origin has fired, we find that replication can be initiated synchronously in the LDDR model for a small range of parameters: At the experimentally measured replication time of \datA of $\tau_{\rm datA}=0.13 \, {\rm h}\approx 8$~min \cite{Kitagawa1996, Katayama2017}, replication is initiated with a high degree of synchrony when the deactivation rate of RIDA becomes high with a delay of $\tau_{\rm rida} = 0.1$~h=6 min after the origin has fired (Fig. \ref{fig:fig_6}b and c). The closer the site \datA is to the origin, the later RIDA should start for synchronous replication initiation (Fig. \ref{fig:fig_6}b). 

It remains however unclear what molecular mechanism could cause a delay in the onset of RIDA of about 6 minutes. In RIDA, the DNA polymerase clamp on newly synthesized DNA forms a complex with ADP and the Hda protein. The resultant ADP-Hda-clamp-DNA can bind ATP-DnaA and stimulates ATP hydrolysis yielding ADP-DnaA \cite{Nakamura2010, Kato2001}. It is conceivable that Hda binding is slow, but whether it would yield a delay of about 6 minutes is far from clear. For experimentally realistic parameters, the LDDR model appears therefore not sufficient to explain how replication is initiated synchronously in \textit{E. coli}.

\section{Titration can enhance the degree of synchrony of an activation switch}

In \textit{E. coli}, DNA replication initiation is not only controlled via an activation switch but also via titration \cite{Atlung1991, Hansen2018}. 
To study the effect of titration on the degree of synchrony, we add homogeneously distributed titration sites on the chromosome to the LDDR model \cite{Berger2022}. In the LDDR-titration model, the initiation potential is given by the free ATP-DnaA concentration $[D]_{\rm ATP,f}$ in the cell and both oscillations in the active fraction $f$ and in the free DnaA concentration $[D]_{\rm T, f}$ contribute to regulating replication initiation. We again model the stochastic opening probability of the origin as a Hill function (equation \ref{eq:opening_prob_f}) with Hill coefficient $m=10$. The critical initiation potential $y^\ast$ is now given by a critical free ATP-DnaA concentration $[D]_{\rm ATP,f}^\ast$ at which ATP-DnaA binds cooperatively to the origin. We here neglect the effect of the relatively small number of about 10-20 DnaA proteins that are bound to the origin on the free DnaA concentration. As explained in Ref. \cite{Berger2022}, we set the parameters (by varying the lipid activation rate $\alpha_{\rm l}$) such that the initiation volume of the switch $v^\ast_{\rm s}$ and the initiation volume of the titration mechanism $v^\ast_{\rm t}$ are approximately the same. This optimal choice ensures that both the free concentration and the active fraction rise at the same critical volume per origin, thus increasing the amplitude of the oscillations in the free ATP-DnaA concentration. 
\begin{figure*}%[tbhp]
	\centering
	\includegraphics[width=0.7\linewidth]{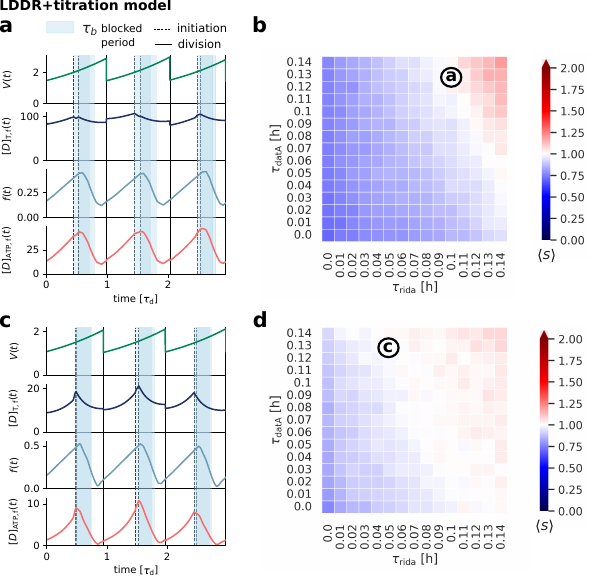}
	\caption{\textbf{Adding titration sites to the LDDR model enhances initiation synchrony for low critical free DnaA concentrations.} (a, c) The volume $V(t)$, free DnaA concentration (independent of whether DnaA is bound to ATP or ADP) $[D]_{\rm T, f}(t)$, the ATP-DnaA fraction $f(t)$, and the free ATP-DnaA concentration $[D]_{\rm ATP, f}(t)$ as a function of time (in units of the doubling time of the cell $\tau_{\rm d}=0.67$~h=40 min) for a critical free ATP-DnaA concentration of $[D]_{\rm f, ATP}^\ast=50 \, \mu {\rm m}^{-3}$ (a) and $[D]_{\rm f, ATP}^\ast=10 \, \mu {\rm m}^{-3}$ (c). During the blocking period $\tau_{\rm b}$ (light blue shaded area), the newly replicated origins cannot be re-initiated. (a) When the critical free ATP-DnaA concentration is relatively high, the free DnaA concentration $[D]_{\rm T, f}(t)$ oscillates only weakly and decreases slightly after new rounds of replication are initiated due to the synthesis of new sites. The shape of the oscillations in the free ATP-DnaA concentration $[D]_{\rm ATP, f}(t)$ is therefore mainly determined by the oscillations in the ATP-DnaA fraction $f(t)$. (b, d) The average degree of synchrony $\langle s \rangle$ as a function of the replication time of the site \datA $\tau_{\rm datA}$ and the onset time of RIDA $\tau_{\rm rida}$ for $[D]_{\rm f, ATP}^\ast=50 \, \mu {\rm m}^{-3}$ (b) and $[D]_{\rm f, ATP}^\ast=10 \, \mu {\rm m}^{-3}$ (d). The sites \textit{DARS1} and \textit{DARS2} are replicated at the experimentally measured times $\tau_{\rm d1}=0.25 \, {\rm h}=15$~min and $\tau_{\rm d1}=0.4 \, {\rm h}=24$~min, respectively. (b) When the critical free ATP-DnaA concentration is high, the effect of the titration sites on the degree of synchrony is small and almost indistinguishable from the scenario without titration sites (compare to Fig. \ref{fig:fig_6}b). (c) At a lower critical free ATP-DnaA concentration, the oscillations in the free concentration are larger and lead to sharper oscillations of the free ATP-DnaA concentration. This causes a broader range of parameters for which replication is initiated synchronously (d). 
		For each parameter pair in b and d, the average degree of synchrony was obtained from $N=5000$ consecutive cell cycles. (See Table \ref{tab:parameters} for all parameters.)}
	\label{fig:fig_7}
\end{figure*}

Fig. \ref{fig:fig_7}a/c show the time traces of the model that combines titration with the activation switch. The small jump in the total free DnaA concentration upon cell division results from the following interplay. Firstly, only one out of two chromosomes is selected per daughter cell (Fig. \ref{fig:fig_7}a/c, second panel). The stochastic firing of the origins causes a temporal delay between the initiation of replication at the respective origins. Moreover, in the growth-rate regime of overlapping replication forks considered here, not all chromosomes have been fully replicated at the moment of cell division. Taken together, this means that at the moment of cell division not all chromosomes have the same number of titration sites (the sites are distributed uniformly). The difference in the number of titration sites per chromosome causes a slight change in the free concentration upon cell division. 

Adding titration sites to the LDDR model affects the degree of synchrony only little when the critical free ATP-DnaA concentration at which replication is initiated is high. When a new round of replication is initiated, new titration sites are generated and the free DnaA concentration drops. As discussed in \cite{Berger2022}, at high growth rates, where multiple chromosomes are present in the cell, new titration sites are however replicated at a similar rate as new DnaA proteins are synthesized. Titration therefore introduces only weak oscillations in the free total DnaA concentration (Fig. \ref{fig:fig_7}a). If the critical free DnaA concentration at which DNA replication is initiated is relatively high, the oscillations in the free DnaA concentration contribute only little to the oscillations in the initiation potential
(Fig. \ref{fig:fig_7}a). In this scenario, adding titration to the LDDR model does not significantly change the degree of synchrony, the optimal position of \datA on the chromosome or the optimal onset time of RIDA (compare Fig. \ref{fig:fig_7}b to Fig. \ref{fig:fig_6}b).

When the free DnaA concentration $[D]_{\rm T, f}$ is however low, titration can significantly enhance the degree of synchrony of the LDDR model. Setting the critical free ATP-DnaA $[D]_{\rm f, ATP}^\ast$ to a value that is comparable to the affinity of the titration sites increases the oscillations in the free DnaA concentration (Fig. \ref{fig:fig_7}c). The resulting sharper rise of the free ATP-DnaA concentration gives rise to a higher degree of synchrony at all positions of \datA and onset times of RIDA (Fig. \ref{fig:fig_7}c). The regime of parameters in which replication is initiated with a high degree of synchrony now extends also to shorter and more realistic onset times of RIDA than in the LDDR model (Fig. \ref{fig:fig_7}d). In summary, the full titration-switch model is able to synchronously initiate replication.

\section{Discussion}
The bacterium \textit{E. coli} initiates replication at several origins synchronously with high precision. How it achieves this high degree of synchrony remained however unknown. In this work, we have revealed several general principles that govern whether replication is initiated synchronously at several origins: (1) the initiation potential must remain high after the first origin has fired so that the remaining origins can fire; (2) origins that have already fired must be prevented from reinitiating immediately as long as not all origins have fired; this necessitates a blocking period and (3) the initiation potential must come down before the blocking period is over to prevent reinitiation of the newly replicated origins. The licensing period, during which the origins can fire, must thus be shorter than the blocking period. The blocking period, in turn, is limited to only 10 minutes \cite{Katayama2010, Lu:1994ee, Waldminghaus:2009em}, which means that the licensing period must be shorter than 10 minutes. To ensure that all origins fire during this short licensing period, the initiation potential must rise sharply, and to guarantee that the initiation potential is low again before the blocking period is over, it also must fall sharply. Synchronous replication initiation thus requires sharp oscillations in the initiation potential. Such oscillations will also give rise to small variations in the initiation volume. Our results therefore predict that the experimentally observed small variation in the initiation volume is a result of the requirement of synchronous replication initiation.

We showed that the previously presented model for the regulation of replication initiation in the bacterium \textit{E. coli} \cite{Berger2022} can ensure a high degree of initiation synchrony for a range of parameters that agree with the experimentally measured ones. We find that if replication initiation is governed by a protein activation switch only, the optimal onset time of the RIDA mechanism would have to be about 6 min in order to ensure synchronous replication initiation. As RIDA is coupled to active replication \cite{Katayama1998}, protein diffusion in cells is typically in the order of seconds rather than minutes \cite{Elf2007} and binding of HdA to the replication clamps is rather strong \cite{Nakamura2010, Kato2001}, it seems natural to assume that the deactivation via RIDA becomes strong directly after a new round of replication starts. It is however conceivable that HdA concentration rises slowly, that HdA binding is slow or that several RIDA complexes are required for a strong deactivation rate of RIDA \cite{Nakamura2010, Kato2001}. Adding a concentration cycle based on titration sites to the activation switch and bringing the system to a regime where the free DnaA concentration is low during the entire cell cycle enhances the degree of synchrony significantly for a broad range of parameters. Importantly, in the combined model replication is initiated with a high degree of synchrony also for shorter onset times of RIDA. Combining an activation cycle with a concentration cycle is therefore likely to be vital to synchronous replication initiation in \textit{E. coli}. 

Increasing the duration of the blocking period would be an easy way for the cell to increase the degree of synchrony. However, also at very fast growth rates, where the doubling time of \textit{E. coli} is about 20 min, the blocking period must remain shorter than the doubling time in order to allow for a new round of replication to start in time. This imposes a natural bound for the duration of the blocking period. Since the duration of the blocking period imposes a hard constraint on synchronous replication initiation, it is tempting to speculate that the requirement of synchronous replication initiation limits the maximal growth rate of \textit{E. coli}.

Also other organisms such as the bacteria \textit{Bacillus subtilis} \cite{Si2017, Sauls2019}, \textit{Mycobacterium smegmatis} \cite{Santi2013} and \textit{Vibrio cholerae}\cite{Egan2004} initiate multiple chromosomes synchronously in certain growth conditions. These bacteria are evolutionarily divergent and have different molecular mechanisms to control the initiation of replication. Nevertheless, the general principles for synchronous replication initiation presented in this work should also remain valid for these organisms.
For example, while the bacterium \textit{B. subtilis} lacks the protein SeqA, it instead contains the protein Spo0A, which can inhibit replication initiation in the \textit{B. subtilis} phage $\phi$29 in vivo and has been shown to bind to specific sites on the origin in vitro \cite{Katayama2010}. These experiments suggest that Spo0A, similar to SeqA in \textit{E. coli}, represses chromosomal replication by binding directly to the origin region of \textit{B. subtilis}. 

Finally, we have not modelled the binding of about 10-20 ATP-DnaA proteins to the origin explicitly. It has however been proposed in the so-called `initiation cascade model' that initiating replication at the first origin could cause other origins to fire as well by releasing the bound initiator proteins into the cytoplasm \cite{Lobner-Olesen1994, Herrick1996}. The resulting higher concentration of free DnaA proteins could lead to a redistribution of the free DnaA proteins to the remaining origins, making the next replication initiation event more likely \cite{Lobner-Olesen1994}. We tested this idea by introducing weak, cooperative origin binding sites to which only ATP-DnaA can bind into our model. When in this extended model the concentration of ATP-DnaA in the cytoplasm rises, the weak binding sites at the respective origins begin to fill up and then trigger the initiation of replication at a randomly selected origin (see Appendix \ref{sec:model_origin_opening_molecular}). After replication has been initiated, the binding sites at the origin that fired become unavailable for binding DnaA for the duration of the blocking period, causing a rise in the free DnaA concentration, as predicted by Ref. \cite{Lobner-Olesen1994}. We find, however, that the ATP-DnaA binding to the origin has two opposing effects: on the one hand, the initiation potential indeed increases right after the first initiation event due to the released ATP-DnaA proteins, making the next initiation event more likely (see Fig. \ref{fig:fig_10_cascade}a and b). On the other hand, binding of ATP-DnaA proteins to the origin leads to a less sharp rise in the free DnaA concentration right before the first origin initiates replication (see Fig. \ref{fig:fig_10_cascade}a and b). A sharp rise of the initiation potential right before replication initiation is however a necessary requirement for synchronous replication initiation. Therefore, the net effect of the initiation cascade on the degree of synchrony is approximately zero and we do not find a significant increase in the degree of synchrony (see Fig. \ref{fig:fig_10_cascade}c).

\section{Acknowledgements}
We want to thank Vahe Galstyan for the fruitful discussions and his mathematical insights, and Lorenzo Olivi for inspiring discussions. We acknowledge financial support from The Netherlands Organization of Scientific Research (NWO/OCW) Gravitation program Building A Synthetic Cell (BaSyC) (024.003.019). 

\counterwithin{figure}{section}
\section{Appendix}
\subsection{coarse-grained model for origin opening}
\label{sec:origin_dynamics_details}
We describe the origin region as a two-state system that can switch between an open (O) or a closed (C) configuration with the opening rate $k_{\rm o}$ and the closing rate $k_{\rm c}$. If the origin is open, replication can be initiated (I) with a maximal firing rate $k_{\rm f}^0$:
\begin{equation}
C \xrightleftharpoons[k{\rm c}]{k_{\rm o}} O \xrightarrow{k_{\rm f}^0} I
\end{equation}
In thermal equilibrium, the ratio of the transition rates between the open and closed state is given by the Boltzman distribution of the energy difference between the two states:
\begin{equation}
\frac{k_{\rm c}}{k_{\rm o}} = \frac{e^{-\beta \, E_{\rm c}}}{e^{-\beta \, E_{\rm o}}} = e^{\beta \, \Delta G}
\end{equation}
with $\beta = k_{\rm B} \, T$ and the energy difference
\begin{equation}
\Delta G = E_{\rm o} - E_{\rm c}
\end{equation}
where $E_{\rm o}$ is the energy of the open state and $E_{\rm c}$ is the energy of the closed state. The probability to be in the open state as a function of the energy difference $\Delta G$ is given by
\begin{equation}
p_{\rm o} = \frac{e^{-\beta \, E_{\rm o}}}{e^{-\beta \, E_{\rm o}} + e^{-\beta \, E_{\rm c}}} = \frac{1}{1 + e^{\beta \, \Delta G}}
\label{eq:opening_prob_energies}
\end{equation}
Assuming rapid opening and closing dynamics of the origin, the origin firing rate is given by equation \ref{eq:firing_rate}.
The higher the initiation potential $f$ in the cell, the more likely is it that the origin is open and that replication can be initiated. We model this observation phenomenologically by assuming that the opening probability $p_{\rm o}$ increases with the activation potential $f$ following a Hill function (see equation \ref{eq:opening_prob_f}).
\begin{table*}%[tbhp]
	\begin{tabularx}{\textwidth}{PGPG}
		Parameter & name & value & Motivation \\
		\hline\hline
		$n$ & Hill coefficient of initiation potential & 5 &  set to match initiation precision reported in \cite{Wallden2015} \\
		$v^\ast$ [$\mu \rm{m}^{3}$] & initiation volume per origin &  1 & set to match initiation volume reported in \cite{Wallden2015} \\
		$m$ & Hill coefficient of opening probability & 10 & \cite{Katayama2017} \\
		$y^\ast$ & critical initiation potential &  0.5 & set to maximal sharpness of opening probability\\
		$K_{\rm D}$ [$\mu \rm{m}^{-3}$] & dissociation constant of (de)activators & 5 &   \cite{Schaper1995} \\
		$\alpha_{\rm l} \, [l]$ [$\mu $m$^{-3}$ \, h$^{-1}$] & activation rate lipids & LDDR:500, LDDR+titration:800 & set to match initiation volume reported in \cite{Wallden2015} \\
		$\beta_{\rm datA}$ [h$^{-1}$] & deactivation rate \textit{datA} & 600 & \cite{Kasho2013} \\ 
		$\tau_{\rm datA}$ [h] & replication time \textit{datA} & 0.13 & \cite{Kasho2013} \\
		$f^\ast$ & critical initiator fraction & 0.5 & \cite{Kurokawa1999, Katayama2001} \\
		$\tau_{\rm i}$ [h] & initiation duration & 0.27 & see Fig. \ref{fig:fig_4_2} \\
		$\alpha_{\rm d1}$ [h$^{-1}$] & activation rate \textit{DARS1} & 100 & \cite{Katayama2017, Kasho2014} \\
		$\tau_{\rm d1}$ [h] & replication time \textit{DARS1} & 0.4 & \cite{Katayama2017} \\
		$\alpha_{\rm d2}^+$ [h$^{-1}$] & high activation rate \textit{DARS2} & 600 & combined with $\beta_{\rm rida}$ \\
		$\alpha_{\rm d2}^-$ [h$^{-1}$] & low activation rate \textit{DARS2} & 50 & set to arbitrary low value \\
		$\tau_{\rm d2}$ [h] & replication time \textit{DARS2} &  0.25 & \cite{Kasho2014} \\
		$\tau_{\rm d2}^+$ [h] & start high activation rate \textit{DARS2} & 0.2 & \cite{Kasho2014} \\
		$\tau_{\rm d2}^-$ [h] & end high activation rate \textit{DARS2} & 2/3 & \cite{Kasho2014} \\
		$\beta_{\rm rida}$ [h$^{-1}$] & deactivation rate RIDA & 500 & \cite{Nakamura2010, Kasho2013, Moolman2014} \\
		$[D]_{\rm T}$ [$\mu \rm{m}^{-3}$] & total DnaA concentration & 400 & \cite{Hansen1991, Speck2001}  \\
		$\phi_{\rm 0}$ & gene allocation fraction &  $4 \times 10^{-4}$ & set to match $[D]_{\rm T}$\\
		$K_{\rm D}^{\rm s}$ [$\mu \rm{m}^{-3}$] & dissociation constant of titration sites & 1 & \cite{Schaper1995} \\
		$n_{\rm ori}^{\rm s}$ & number of origin binding sites & 10 & \cite{Katayama2017} \\
		$[D]_{\rm ATP,f}^{\ast}$ & critical free ATP-DnaA concentration & 10 & \cite{Schaper1995} \\
		$\rho$ [$\mu \rm{m}^{-3}$] & number density & $10^6$ & \cite{Milo2013} \\
		$k_{\rm f}^0$ [$\rm{s}^{-1}$] & maximal origin firing rate & 1000 & set such that degree of synchrony is maximal \\
		$\tau_{\rm b}$ [h] & blocking period & 0.17 & \cite{Campbell:1990it,Lu:1994ee, Waldminghaus:2009em} \\
		$\lambda$ [h$^{-1}$] & growth rate & 1.04 & \cite{Si2017, Wallden2016} \\
		$T_{\rm C}$ [h] & C-period & 2/3 & \cite{Cooper1968} \\
		$T_{\rm D}$ [h] & D-period & 1/3 & \cite{Cooper1968} \\

		\hline\hline
	\end{tabularx}
	\caption
	{\label{tab:parameters}
	{\bf Parameters used in the simulations} One molecule per cubic micrometer corresponds to approximately one nM ($1~\mu \rm{m}^{-3}= 1.67$~nM).
	}
\end{table*}

\subsection{Derivation of approximation for opening probability}
\label{sec:methods_deriv_eff_hill_coeff}

\begin{figure*}%[tbhp]
	\centering
	\includegraphics[width=0.65\linewidth]{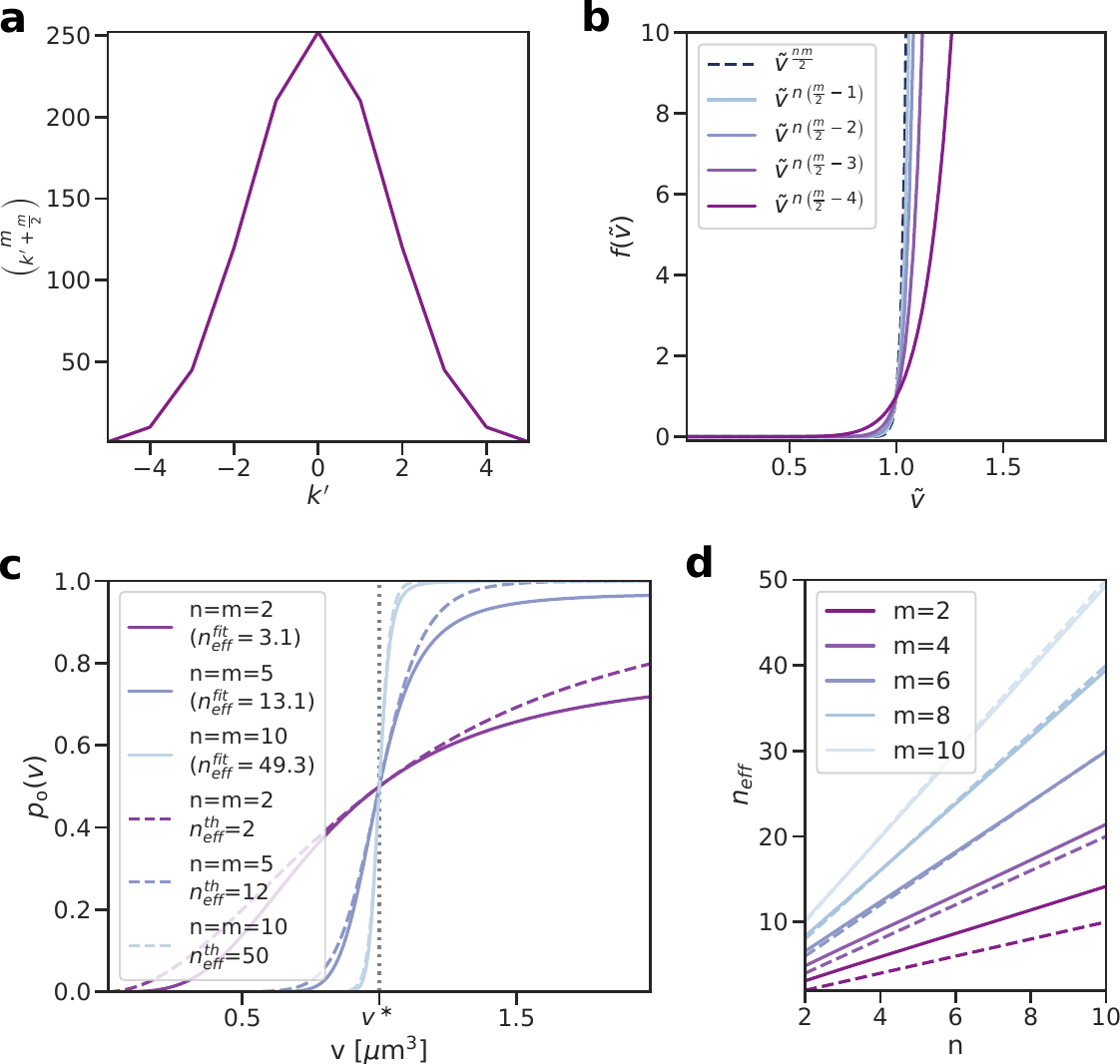}
	\caption{\textbf{The instantaneous opening probability can be approximated by a Hill function with an effective Hill coefficient} (a) The binomial coefficient as defined in equation \ref{eq:binomial_coeff_def} as a function of the index $k^\prime$ for $m=10$. The binomial coefficient is centered around and maximal at $k^\prime = 0$ and becomes small for $k^\prime \gg 0$. (b) The second term in equation \ref{eq:geometric_series_shifted} for different values of the index $k^\prime$ as a function of the rescaled volume $\tilde{v}= v/v^\ast$. For small values of $k^\prime$, the second term in equation \ref{eq:geometric_series_shifted} is well approximated by the term $\tilde{v}^{\frac{n \, m}{2}}$ (dashed blue line). (c) The opening probability of the origin $p_{\rm o}(f(v))$ (equation \ref{eq:opening_prob_of_v}) as a function of the volume per origin $v$ for different Hill coefficients $n$ and $m$ (solid lines).  The effective Hill coefficient $n_{\rm eff}^{\rm fit}$ is obtained from a fit of the function $p_{\rm o}(f(v))$ to a Hill function (equation \ref{eq:opening_prob_of_v_fitted}). The dashed lines show the approximated opening probability (equation \ref{eq:opening_prob_of_v_approx_final}) with the effective Hill coefficient as defined in equation \ref{eq:n_eff_approx_definition}. The vertical dotted line indicates the critical volume per origin $v^\ast$ at which the opening probability equals 1/2. (d) The fitted (solid line) and the approximated (dashed line, equation \ref{eq:n_eff_approx_definition}) Hill coefficient as a function of the Hill coefficient $n$ for different values of the Hill coefficient $m$. Except for very low Hill coefficients $n$ and $m$, the approximated Hill coefficient agree well. In all graphs $f^\ast=v^\ast= 0.5$.(See Table \ref{tab:parameters} for all parameters.)}
	\label{fig:fig_A1}
\end{figure*}
We want to find an analytical expression for the opening probability $p_{\rm o}$ and therefore also the instantaneous firing rate $k_{\rm f}$ (equation \ref{eq:firing_rate}) as a function of time. We therefore insert equation \ref{eq:initiation_potential} into equation \ref{eq:opening_prob_f} to obtain:
\begin{align}
p_{\rm o}(f(v))=& \frac{v^{n \, m}}{f^{\ast \, m} \, (v^{\ast \, n}+ v^{ n})^{ m}+ v^{n \, m}} \\
=& \frac{v^{n \, m}}{f^{\ast \, m} \, v^{\ast \, n \, m} \, (1+ \tilde{v}^{n})^{m}+ v^{n \, m}},
\label{eq:opening_prob_of_v}
\end{align}
where we used $\tilde{v}=v/v^\ast$.  According to the binomial formula, we can write
\begin{align}
(1+\tilde{v}^{ n})^{m} &= \sum_{k=0}^{m} \binom{m}{k} \, 1^k \, (\tilde{v}^n)^{m-k} \\ &
=\sum_{k=0}^{m} \binom{m}{k} \,(\tilde{v}^n)^{m-k}.
\label{eq:geometric_series}
\end{align}
with the binomial coefficient
\begin{equation}
\binom{m}{k} \coloneqq \frac{m!}{k! \, (m-k)!}.
\label{eq:binomial_coeff_def}
\end{equation}
We introduce the shifted parameter $k^\prime = k - m/2$, such that equation \ref{eq:geometric_series} can be rewritten as:
\begin{equation}
(1+\tilde{v}^{ n})^{m} = \sum_{k^\prime=-m/2}^{m/2} \binom{m}{k^\prime + \frac{m}{2}} \,  (\tilde{v}^n)^{\frac{m}{2}-k^\prime}.
\label{eq:geometric_series_shifted}
\end{equation}
By examining the first and the second term of the sum in equation \ref{eq:geometric_series_shifted} separately, we find that the binomial coefficient has a maximum at $k^\prime = 0$ and decays quickly for $k^\prime \neq 0$ (See Fig. \ref{fig:fig_A1}a). Secondly, as can be seen in Figure  \ref{fig:fig_A1}b, for small  $k^\prime \ll \pm m/2$ and sufficiently large Hill coefficient $m$, the second term is approximately given by
\begin{equation}
\tilde{v}^{ n \, (\frac{m}{2}-k^\prime)} \approx  \tilde{v}^\frac{m n}{2}.
\end{equation}
Combining these two observations, we can approximate equation \ref{eq:geometric_series_shifted} by
\begin{equation}
(1+\tilde{v}^{ n})^{m} \approx \sum_{k^\prime=-m/2}^{m/2} \binom{m}{k^\prime + \frac{m}{2}} \,  \tilde{v}^\frac{m n}{2}.
\end{equation}
Finally, using that 
\begin{equation}
\sum_{k^\prime=-m/2}^{m/2} \binom{m}{k^\prime + \frac{m}{2}} = 2^m,
\end{equation}
we find 
\begin{equation}
(1+\tilde{v}^{ n})^{m} \approx\, 2^m \, \tilde{v}^\frac{m n}{2}
\end{equation}
Plugging this expression into equation \ref{eq:opening_prob_of_v} gives
\begin{align}
p_{\rm o}(v)\approx& \frac{v^{n \, m}}{f^{\ast \, m} \, v^{\ast \, n \, m} \,  2^m \, \tilde{v}^\frac{m n}{2}+ v^{n \, m}}\\
=& \frac{v^{n \, m}}{f^{\ast \, m}\,  2^m  \, v^{\ast \, \frac{n m}{2}} \, v^\frac{m n}{2}+ v^{n \, m}}\\
=& \frac{v^{\frac{n m}{2}}}{f^{\ast \, m}\,  2^m  \, v^{\ast \, \frac{n m}{2}}+ v^{\frac{n m}{2}}}
\label{eq:opening_prob_of_v_approx}
\end{align}
For $f^{\ast}=0.5$ we then find equation \ref{eq:opening_prob_of_v_approx_final} of the main text with the effective Hill coefficient $n_{\rm eff} = n \, m /2$.
By comparing the approximation of $p_{\rm o}(v)$ in equation \ref{eq:opening_prob_of_v_approx_final} to a function 
\begin{equation}
p_{\rm o}^{\rm fit}(v) = a^{\rm fit} \, \frac{v^{n_{\rm eff}^{\rm fit}}}{v^{\ast \, n_{\rm eff}^{\rm fit}}+ v^{n_{\rm eff}^{\rm fit}}}
\label{eq:opening_prob_of_v_fitted}
\end{equation}
that was fitted to $p_{\rm o}(f(v))$ (equation \ref{eq:opening_prob_of_v}), we find that the approximation in equation \ref{eq:opening_prob_of_v_approx_final} 
is indeed a good approximation for sufficiently large Hill coefficients $n$ and $m$, especially at volume close to the critical volume $v^\ast$ (Fig. \ref{fig:fig_A1}c). Indeed, the fitted Hill coefficient agrees well with the approximated Hill coefficient in equation \ref{eq:n_eff_approx_definition} for a broad range of Hill coefficients $n$ and $m$, respectively (Fig. \ref{fig:fig_A1}d).

\subsection{Parameter choice for maximal firing rate}
\label{sec:parameter_choice_max_opening_rate}
Combining the approximation for the opening probability as a function of the volume per origin (equation \ref{eq:opening_prob_of_v_approx_final}), the exponentially growing cell-volume $V(t)=V_{\rm b} \, e^{\lambda \, t}$, and the expression for the firing rate (equation \ref{eq:firing_rate}), we find the following time-dependent firing rate of a single origin:
\begin{equation}
k_{\rm f}(t)= k_{\rm f}^0 \frac{\left(V_{\rm b} \, e^{\lambda \, t}\right)^{n_{\rm eff}}}{v^{\ast \, n_{\rm eff}}+ \left(V_{\rm b} \, e^{\lambda \, t}\right)^{n_{\rm eff}}}
\label{eq:firing_rate_approx}
\end{equation}
From this rate, we can calculate the survival probability 
\begin{align}
S(t)=& e^{- \int_{t_{0}}^{t} dt^\prime k_{\rm f}(t^\prime)} \\
=& e^{- \frac{k_{\rm f}^0}{n_{\rm eff} \, \lambda} \, \ln\left(\frac{(V_{\rm b} \, e^{\lambda \, t})^{n_{\rm eff}} + v^{\ast \, n_{\rm eff}}}{V_{\rm b}^{n_{\rm eff}} + v^{\ast \, n_{\rm eff}}}\right)}
\end{align}
where we solved the integral with the initial condition $S(t_0=0)=1$. We now impose that at the theoretical initiation volume per origin $v(t=t^\ast)=v^\ast$, the survival probability is exactly $S(t^\ast)=0.5$. Using this constraint, we obtain the following expression for the maximal firing rate as a function of the effective Hill coefficient $n_{\rm eff}$:
\begin{equation}
k_{\rm f}^0 (n_{\rm eff}) = \frac{n_{\rm eff} \, \lambda \, \ln(2)}{\ln\left(\frac{2 \, v^{\ast n_{\rm eff}}}{V_{\rm b}^{n_{\rm eff}} + v^{\ast n_{\rm eff}}} \right)}
\label{eq:parameter_choice_firing_rate}
\end{equation}
This parameter choice ensures that the average initiation volume $\langle v^\ast \rangle$ is given by $v^\ast$.

\subsection{Derivation of the theoretical prediction for the degree of synchrony}
\label{sec:derivation_degree_synch_th}
In the following, we derive a theoretical prediction for the probability that two initiation events happen within a time interval $\tau_{l}$. We assume here that the two firing events are statistically independent, meaning that between the first initiation event at time $t_1$ and the time $t_1 +\tau_{l}$, the change in the number of origins induced by the first event has no effect on the firing of the second event.
%Correspondingly, the waiting time distribution is given by 
%\begin{equation}
%w(t)=  -\frac{\partial S}{\partial t} = k_{\rm f}(t) \, e^{- \int_{t_{0}}^{t} dt^\prime k_{\rm f}(t^\prime)}
%\end{equation}
Using the firing rate in equation \ref{eq:firing_rate_approx}, we can now calculate the error probability $S_{\rm err}$ that the second event does \textit{not} happen within a time $\tau_{\rm l}$ after the first event, given that the first event happened at time $t_1$:
\begin{align}
S_{\rm err}(t_2-t_1 >\tau_{\rm l}|t_1) &=e^{- \int_{t_1}^{t_1 + \tau_{\rm l}} dt^\prime k_{\rm f}(t^\prime)} \\
&= e^{- \frac{k_{\rm f}^0}{n_{\rm eff} \, \lambda} \, \log\left(\frac{\left(V_{\rm b} \, e^{\lambda \, \left(t_1 + \tau_{\rm l}\right)}\right)^{n_{\rm eff}} + v^{\ast \, n_{\rm eff}}}{\left(V_{\rm b} \, e^{\lambda \, t_1}\right)^{n_{\rm eff}} + v^{\ast \, n_{\rm eff}}}\right)}
\end{align}
The average error probability $\langle S_{\rm err} \rangle$ over all initiation times of the first event, $t_1$, is then given by 
\begin{align}
\langle S_{\rm err} \rangle &= \int_{0}^{\tau_{\rm d}} dt_1 \, q_1(t_1) \, S_{\rm err}( t_2 - t_1 >\tau_{\rm l}|t_1) \\
&=  \int_{0}^{\tau_{\rm d}} dt_1 \, q_1(t_1) \, e^{- \int_{t_1}^{t_1 + \tau_{\rm l}} dt^\prime k_{\rm f}(t^\prime)}
\label{eq:average_err_prob}
\end{align}
where $\tau_{\rm d}$ is the doubling time of the cell. The propensity $q_1(t_1)$ that one out of two origin events happens at time $t_1$ is given by 
\begin{equation}
q_1(t_1) = 2 \, k_{\rm f}(t_1) \, e^{- \int_{t_{0}}^{t_1} dt^\prime 2 \, k_{\rm f}(t^\prime)}
\label{eq:propensity_q1}
\end{equation}
Therefore, the average error probability $\langle S_{\rm err} \rangle$ that the second origin fires after a time $\tau_{\rm l}$ after the first event is given by plugging expression \ref{eq:propensity_q1} into equation \ref{eq:average_err_prob}:
\begin{align}
\langle S_{\rm err} \rangle &=  2 \, \int_{0}^{\tau_{\rm d}} dt_1 \, k_{\rm f}(t_1) \, e^{- 2 \, \int_{t_{0}}^{t_1} dt^\prime k_{\rm f}(t^\prime)} \, e^{- \int_{t_1}^{t_1 + \tau_{\rm l}} dt^\prime k_{\rm f}(t^\prime)} \\
\nonumber
&= 2 \, k_{\rm f}^0 \int_{0}^{\tau_{\rm d}} dt_1 \, \frac{\left(V_{\rm b} \, e^{\lambda \, t_1}\right)^{n_{\rm eff}}}{v^{\ast \, n_{\rm eff}}+ \left(V_{\rm b} \, e^{\lambda \, t_1}\right)^{n_{\rm eff}}} \\
&\qquad \qquad  \times \, e^{- \frac{2 \, k_{\rm f}^0}{n_{\rm eff} \, \lambda} \, \ln\left(\frac{\left(V_{\rm b} \, e^{\lambda \, t_1} \right)^{n_{\rm eff}} + v^{\ast \, n_{\rm eff}}}{V_{\rm b}^{n_{\rm eff}} + v^{\ast \, n_{\rm eff}}}\right)} \nonumber \\
&\qquad \qquad  \times \,\, e^{- \frac{k_{\rm f}^0}{n_{\rm eff} \, \lambda} \, \ln\left(\frac{\left(V_{\rm b} \, e^{\lambda \, (t_1 + \tau_l)} \right)^{n_{\rm eff}} + v^{\ast \, n_{\rm eff}}}{\left(V_{\rm b} \, e^{\lambda \, t_1} \right)^{n_{\rm eff}} + v^{\ast \, n_{\rm eff}}}\right)},
\label{eq:error_prob_int}
\end{align}
where $\tau_{\rm d}$ is the average division time of the cell and $\tau_{\rm d}\gg \tau_{\rm l}$, such that the probability that both origins have not yet fired at $\tau_{\rm d}$ becomes negligible. The average probability that the second origin fires within a time interval $\Delta t = t_2 - t_1< \tau_{\rm l}$ after the first has fired at $t_1$,  is then given by:
\begin{equation}
\langle P(\Delta t < \tau_{\rm l})\rangle = 1-\langle S_{\rm err} (\tau_{\rm l}) \rangle
\end{equation}
We solve the integral in equation \ref{eq:error_prob_int} numerically and use expression \ref{eq:degree_synchr_th} to predict the degree of synchrony for two origins (see Fig. \ref{fig:fig_3_synch}b). 

One can also calculate analytically the degree of synchrony at higher growth rates where there are typically four or more origins per cell at the beginning of an initiation cascade. The probability that none of the $n-1$ origins fire within the time $\tau_{\rm l}$ after the first origin has fired at $t_1$ is similar to equation \ref{eq:error_prob_int} and given by
\begin{equation}
	\langle S_{\rm err} \rangle =  n \, \int_{0}^{\tau_{\rm d}} dt_1 \, k_{\rm f}(t_1) \, e^{- n \, \int_{t_{0}}^{t_1} dt^\prime k_{\rm f}(t^\prime)} \, e^{- (n-1) \int_{t_1}^{t_1 + \tau_{\rm l}} dt^\prime k_{\rm f}(t^\prime)}
\end{equation}
This is the probability that given that the first origin fires at $t_1$, all $n-1$ other origins fire later than $t_1 + \tau_{\rm l}$. Importantly, one now also needs to take into account the cases where only one or more origins fire $t_1 + \tau_{\rm l}$ and the other fire before. We here do not derive an expression for the scenario $n>2$. 

\subsection{Derivation of theoretical prediction for $\langle \Delta t \rangle$ and the CV of the initiation volume}
\label{sec:deriv_average_initiation_interval_cv}
The average time interval between two independent firing events $\langle \Delta t \rangle$ can be calculated analytically for the approximate opening probability in equation \ref{eq:opening_prob_of_v_approx_final} via
\begin{align}
\langle \Delta t \rangle =&  \int_{0}^{\tau_{\rm d}} dt_1 \, \int_{t_1}^{\tau_{\rm d}} dt_2 \, 2 \,k_{\rm f}(t_1) \, k_{\rm f}(t_2) \,e^{- 2 \, \int_{t_{0}}^{t_1} dt^\prime k_{\rm f}(t^\prime)} \nonumber\\
& \times  e^{- \int_{t_1}^{t_2} dt^{\prime\prime} k_{\rm f}(t^{\prime\prime})}
\end{align}
Solving this integral numerically gives the pink line in Fig. \ref{fig:fig_3_synch}b. 

The theoretical coefficient of variation of the initiation volume $V$ is given by
\begin{equation}
CV=\frac{\sigma}{\mu}= \frac{\sqrt{\langle V^2 \rangle - \langle V \rangle^2}}{\langle V \rangle}
\label{eq:CV_definition}
\end{equation}
where we use
\begin{equation}
\langle V \rangle = \int_{0}^{\tau_{\rm d}} dt \,k_{\rm f}(t) \, e^{- \int_{t_{0}}^{t} dt^\prime k_{\rm f}(t^\prime)} \, V_{\rm b} \, e^{\lambda \, t}
\end{equation}
and 
\begin{equation}
\langle V^2 \rangle = \int_{0}^{\tau_{\rm d}} dt \,k_{\rm f}(t) \, e^{- \int_{t_{0}}^{t} dt^\prime k_{\rm f}(t^\prime)} \, \left( V_{\rm b} \, e^{\lambda \, t} \right)^2
\end{equation}
In this theoretical model, the only source of noise is intrinsic noise and the CV in equation \ref{eq:CV_definition} therefore also corresponds to the intrinsic noise as defined by Ref. \cite{Boesen2022} based on the derivation of Elowitz et al. \cite{Elowitz2002}.

\begin{figure*}%[tbhp]
	\centering
	\includegraphics[width=0.5\linewidth]{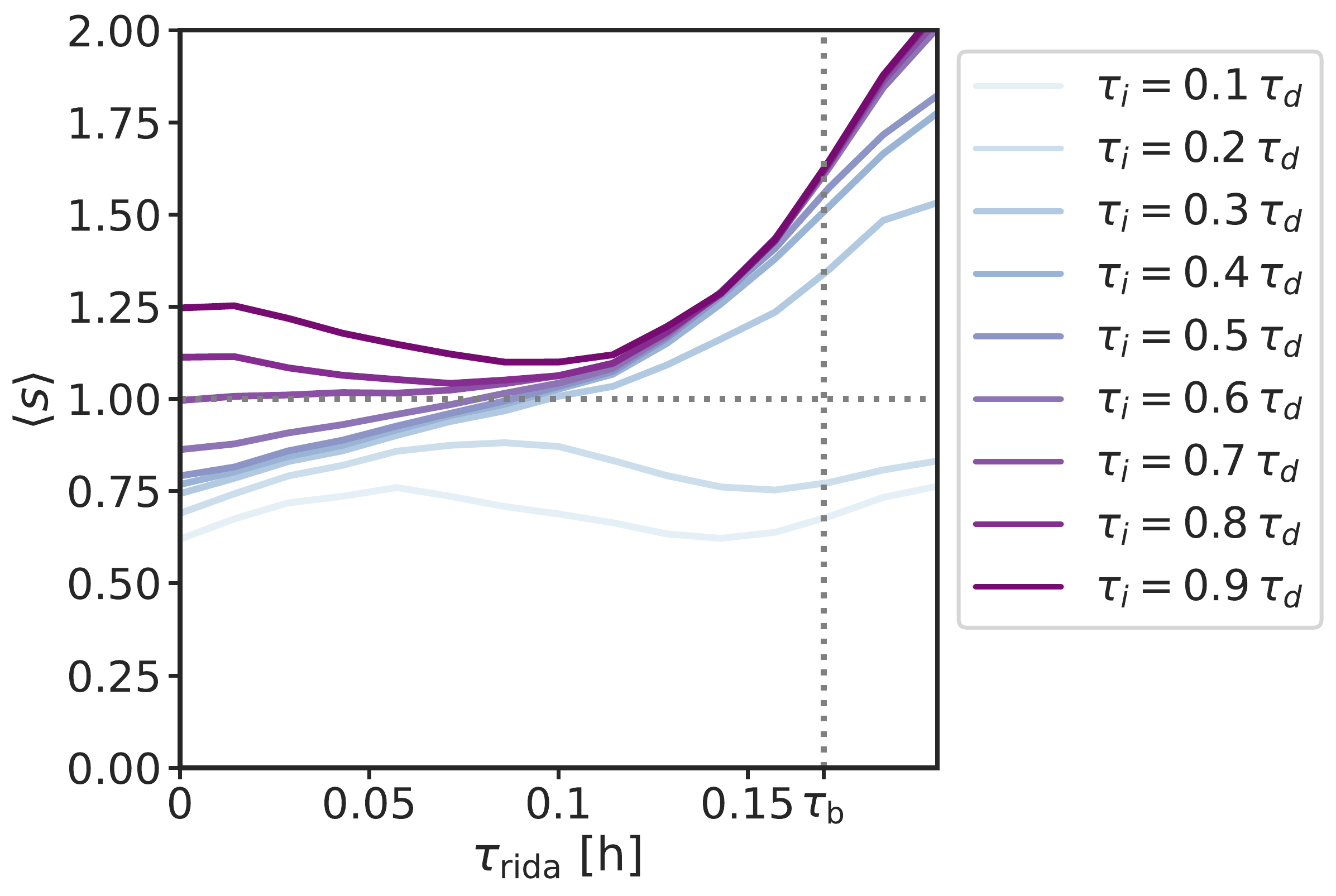}
	\caption{\textbf{The average degree of synchrony $\langle s \rangle$ depends strongly on the initiation duration when replication is initiated asynchronously, but not when replication is initiated synchronously.} The average degree of synchrony $\langle s \rangle$ as a function of the starting time of RIDA $\tau_{\rm rida}$ for varying initiation durations $\tau_{\rm i}$ (in units of the doubling time $\tau_{\rm d}$) for the LDDR model. The degree of synchrony $s$ is obtained by counting the number of origin firing events from the first origin firing until the end of the initiation duration $\tau_{\rm i}$. When replication is initiated synchronously at all origins within a short time interval (at $\tau_{\rm rida} \approx 0.1$ h), the average degree of synchrony does not depend strongly on the initiation period $\tau_{\rm i}$. When origins are however initiated asynchronously over the course of the cell cycle, the average degree of synchrony can either be smaller or larger than one, depending on the duration $\tau_{\rm i}$. In the rest of this manuscript, we use an intermediate initiation duration of $\tau_{\rm i}= 0.4 \, \tau_{\rm d}$. (See Table \ref{tab:parameters} for all parameters.)}
	\label{fig:fig_4_2}
\end{figure*}
\subsection{Definition of the initiation duration in LDDR and LDDR-titration models}
\label{sec:definition_cascade_duration}

While a synchronization parameter cannot be defined uniquely, we will define one to quantify the degree to which replication is initiated synchronously and then show that the result is fairly robust to the precise definition. Specifically, the degree of synchrony is obtained by counting the number of origin firing events per initiation event, where the initiation duration $\tau_{ \rm i}$ is a parameter that we will choose carefully (Fig. \ref{fig:fig_2_2_synch}a). In the coarse-grained model, an initiation event starts when the first origin initiates and ends after the licensing period is over. As after the end of the licensing period the initiation potential drops instantaneously to a very low value, re-initiation events after the end of the licensing period are very unlikely in the coarse-grained model. In the LDDR model, the active fraction $f$ does however not decrease instantaneously after RIDA has started and the site \datA has been doubled (Fig. \ref{fig:fig_6}c). Therefore, it is less clear what the initiation period should be. We test the effect of varying the initiation duration $\tau_{ \rm i}$ on the average degree of synchrony $\langle s \rangle$ for different starting times of RIDA  $\tau_{\rm rida}$ (Fig. \ref{fig:fig_6}). The average degree of synchrony $\langle s \rangle$ varies strongly with the initiation duration in parameter regimes where replication is initiated asynchronously: At very low starting times of RIDA  $\tau_{\rm rida}$ replication is under-initiated (Fig. \ref{fig:fig_6}b), but the degree of synchrony nevertheless becomes larger than one at high initiation durations $\tau_{\rm i}> 0.6 \, \tau_{\rm d}$ (Fig. \ref{fig:fig_4_2}). The larger the initiation duration $\tau_{ \rm i}$ the more origin firing events are counted per initiation event, leading to an average degree of synchrony that is larger than one. Conversely, when the RIDA is starting too late and replication is over-initiated (Fig. \ref{fig:fig_6}b), the degree of synchrony can nevertheless be smaller than one if the initiation duration is chosen too short. At the optimal starting time of RIDA of $\tau_{\rm rida}=0.1 \, {\rm h} \approx 6$~min, where replication is initiated synchronously, the choice of the initiation duration becomes however less relevant: Because all origin firing events happen within a relatively small time window, increasing the initiation duration further does not change the degree of synchrony significantly. Only if the initiation duration is chosen way too small ($\tau_{\rm i}=0.2 \, \tau_{\rm d}\approx8$~min) or too large  ($\tau_{\rm i}=0.9 \, \tau_{\rm d}\approx36$~min) becomes the average degree of synchrony smaller or larger than one. We therefore in the following choose an intermediate initiation duration of $\tau_{\rm i}=0.4 \, \tau_{\rm d}\approx16$~min.

\subsection{Model for the initiation cascade}
\label{sec:model_origin_opening_molecular}

\begin{figure*}%[tbhp]
	\centering
	\includegraphics[width=0.6\linewidth]{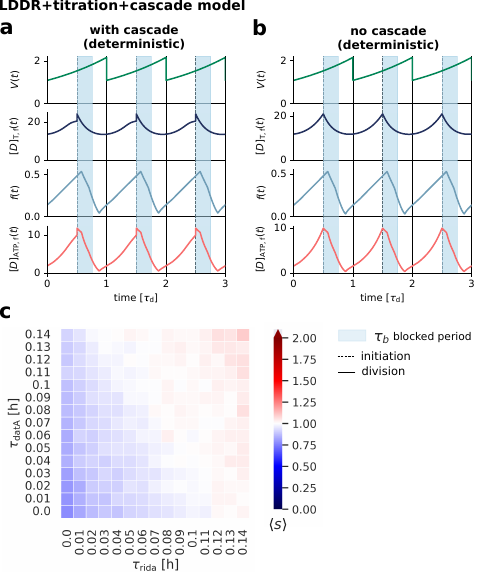}
	\caption{\textbf{The initiation cascade does not significantly enhance the degree of synchrony.} (a, b) The volume $V(t)$, free DnaA concentration (independent of whether DnaA is bound to ATP or ADP) $[D]_{\rm T, f}(t)$, the ATP-DnaA fraction $f(t)$, and the free ATP-DnaA concentration $[D]_{\rm ATP, f}(t)$ as a function of time (in units of the doubling time of the cell $\tau_{\rm d}=0.67$~h=40 min) for a critical free ATP-DnaA concentration of $[D]_{\rm f, ATP}^\ast=10 \, \mu {\rm m}^{-3}$ with (a) and without (b) the initiation cascade. For illustration purposes, replication is here initiated deterministically at all origins as soon as the critical free ATP-DnaA concentration in the cell is reached. (a) When the free ATP-DnaA concentration approaches the critical free ATP-DnaA concentration of $[D]_{\rm f, ATP}^\ast$, ATP-DnaA proteins begin to bind to the weak, cooperative origin binding sites. This leads to a decrease in the rise of the free DnaA concentration right before replication initiation. Upon replication initiation, the bound ATP-DnaA proteins become unavailable, leading to a sharp increase in the free DnaA concentration. (b) For comparison, we here also show the time traces of a system in which the origin binding sites are not modelled explicitly and the only binding sites are the strong DnaA boxes distributed homogeneously all over the chromosome. (c) The average degree of synchrony $\langle s \rangle$ as a function of the replication time of the site \datA $\tau_{\rm datA}$ and the onset time of RIDA $\tau_{\rm rida}$ for the same parameters as in Fig. \ref{fig:fig_7}d. Comparing the two panels reveals that the initiation cascade does not significantly enhance the degree of synchrony. For each parameter pair in c, the average degree of synchrony was obtained from $N=1000$ consecutive cell cycles. (See Table \ref{tab:parameters} for all parameters.)}
	\label{fig:fig_10_cascade}
\end{figure*}
To initiate DNA replication, 8 ATP-DnaA proteins and 3 DnaA proteins independent of the nucleotide-binding state form a cooperative complex at the origin, which induces a conformational change and leads to the opening of the origin \cite{Katayama2017}. Consequently, the DNA replication machinery binds to the open origin, and replication is initiated.
Upon replication initiation, the DnaA proteins that were bound to the origin are likely being released into the cytoplasm, leading to a transient rise in the free DnaA concentration. It has been suggested that the release of origin-bound DnaA proteins from one origin triggers replication initiation at the remaining origins in a so-called `initiation cascade' \cite{Lobner-Olesen1994}. Here we propose a model to test whether the rise in the free DnaA concentration upon replication initiation at one origin could be sufficient to trigger replication initiation at the other origins.

So far, we have modelled the origin opening and firing process in a coarse-grained manner using a Hill function as a function of the initiation potential for the opening probability of the origin (see Appendix \ref{sec:origin_dynamics_details}). Now, we instead model the binding of ATP-DnaA to the origin explicitly by introducing weak, cooperative binding sites for ATP-DnaA proteins at the origin. Specifically, we neglect the three strong binding sites to which both ATP- and ADP-DnaA can bind and assume that there are $n$ weak binding sites with the dissociation constant $K_{\rm D}^{\rm ori}$ to which only ATP-DnaA can bind cooperatively. The probability that $n$ ATP-DnaA proteins are bound to the origin is given by 
\begin{equation}
p_{\rm b}^{\rm n}= \frac{Z_{\rm b}^{\rm n}}{\sum_{i=0}^{N}Z_{\rm i}}
\end{equation}
where $Z_{\rm b}^{\rm n}$ is the partition function of $n$ proteins bound to the origin and $\sum_{i=0}^{N}Z_{\rm i}$ is the sum over all possible configurations the origin can be in. Let us first consider the scenario of only two cooperative binding sites. This gives rise to the following probability that two ATP-DnaA proteins are bound to the origin:
\begin{equation}
p_{\rm b}^{\rm 2}= \frac{Z_{\rm b}^{\rm 2}}{Z_{\rm b}^{\rm 0}+2 \, Z_{\rm b}^{\rm 1}+ Z_{\rm b}^{\rm 2}}
\end{equation}
The statistical weight of zero bound ATP-DnaA proteins is normalized to one $Z_{\rm b}^{\rm 0}=1$ and the weight of one bound ATP-DnaA protein is given by $Z_{\rm b}^{\rm 1}=[D]_{\rm f, ATP}/K_{\rm D}$ with the dissociation constant $K_{\rm D}= c_0^{-1} \, e^{-\beta \, \Delta G}$ and the free ATP-DnaA concentration $[D]_{\rm f, ATP}$. The weight of two bound ATP-DnaA proteins is then given by $Z_{\rm b}^{\rm 2}=w \, [D]_{\rm f, ATP}^2/K_{\rm D}^2$ where $w=e^{\beta \, \Delta E}$ accounts for the additional energy gain from cooperative binding of two ATP-DnaA proteins. When cooperative binding is very strong then $\Delta E \gg \Delta G$ and we can neglect terms with lower powers of $w$:
\begin{equation}
p_{\rm b}^{\rm 2}\approx \frac{Z_{\rm b}^{\rm 2}}{Z_{\rm b}^{\rm 0}+ Z_{\rm b}^{\rm 2}}= \frac{w \, [D]_{\rm f, ATP}^2/K_{\rm D}^2}{1+ w \, [D]_{\rm f, ATP}^2/K_{\rm D}^2}=  \frac{[D]_{\rm f, ATP}^2}{\left(\frac{K_{\rm D}}{\sqrt{w}}\right)^2+ [D]_{\rm f, ATP}^2}
\end{equation}
This expression can be generalized to the case of $n$ strongly cooperative ATP-DnaA origin binding sites:
\begin{equation}
p_{\rm b}^{\rm n}\approx\frac{[D]_{\rm f, ATP}^n}{\left(\frac{K_{\rm D}}{\sqrt[n]{w}}\right)^n+ [D]_{\rm f, ATP}^n}
\end{equation}
We therefore recover the expression \ref{eq:opening_prob_f} for the origin opening probability in the coarse-grained model where now the critical free ATP-DnaA concentration is given by $[D]_{\rm ATP,f}^\ast=K_{\rm D}/\sqrt[n]{w}$. 

In order to calculate the free DnaA concentration $[D]_{\rm f}$ in the scenario where both DnaA forms can bind to the 300 homogeneously distributed strong binding sites on the chromosome and ATP-DnaA can additionally bind cooperatively to $n$ weak binding sites on the origin, we write down the following expression
\begin{equation}
[D]_{\rm f} = [D]_{\rm T} - [D]_{\rm s} - [D]_{\rm o} 
\end{equation}
where $[D]_{\rm T}$ is the total DnaA concentration in the cell, $[D]_{\rm s}$ is the concentration of titration-site bound DnaA and $[D]_{\rm o}$ is the origin-bound concentration of ATP-DnaA. An expression for the titration-site bound concentration $[D]_{\rm s}$ as a function of the free DnaA concentration $[D]_{\rm f}$ is obtained from the quasi-equilibrium approximation as explained in \cite{Berger2022}
\begin{equation}
[D]_{\rm s} = \frac{[s]_{\rm T}\, [D]_{\rm f}}{K_{\rm D}^{\rm s} + [D]_{\rm f}}
\end{equation}
with the total titration site concentration $[s]_{\rm T}$ and the dissociation concentration of the titration sites $K_{\rm D}^{\rm s}$. The origin-bound ATP-DnaA concentration $[D]_{\rm o} $ is given by the probability $p_{\rm b}^{\rm n}$ that $n$ ATP-DnaA proteins are bound to the origin times the total concentration of proteins that can be bound to these origin sites. This total concentration is given by the concentration of origins that are available for ATP-DnaA binding $[n_{\rm ori}^{\rm f}]$ times the number of binding sites per origin $n$. Therefore, we obtain the following expression for the free DnaA concentration:
\begin{equation}
[D]_{\rm f} = [D]_{\rm T} - \frac{[s]_{\rm T}\, [D]_{\rm f}}{K_{\rm D}^{\rm s} + [D]_{\rm f}} - \frac{n \, [n_{\rm ori}^{\rm f}] \left([D]_{\rm f} \, f \right)^n}{\left(\frac{K_{\rm D}}{\sqrt[n]{w}}\right)^n+ \left([D]_{\rm f} \, f \right)^n}
\label{eq:free_conc_solved_ori_sites}
\end{equation}
We here made the simplifying assumption that the free ATP-DnaA concentration $[D]_{\rm f, ATP}$ is given by the ATP-DnaA fraction $f$ times the free DnaA concentration $[D]_{\rm f}$. This is a reasonable approximation because the number of origin binding sites is small compared to the total number of DnaA proteins and the total number of titration sites. The total fraction of ATP-DnaA proteins in the cell $f$ is therefore approximately equal to the fraction of ATP-DnaA proteins in the cytoplasm. Importantly, as explained in \cite{Berger2022}, we assume that the switch components (de)activate DnaA independent of whether it is bound to the chromosome (either titration sites or origin sites) or freely diffusing in the cytoplasm. We solve equation \ref{eq:free_conc_solved_ori_sites} numerically at every time step of the simulations given the total titration site concentration $[s]_{\rm T}$, the total DnaA concentration $[D]_{\rm T}$ and the total number of for ATP-DnaA available origin binding sites $[n_{\rm ori}^{\rm f}]$. Replication is again initiated stochastically at every origin with a rate $k_{\rm f}= k_{\rm f}^0 \, p_{\rm b}^{\rm n}$. We model the effect that DnaA proteins are released to the cytoplasm upon replication initiation by transiently reducing the number of available origin binding sites for a duration of $\tau_{\rm b}=10$~min after an origin has initiated replication.

Modelling the ATP-DnaA binding to the origins explicitly does not significantly increase the degree of synchrony for a broad range of parameters. 
When the free ATP-DnaA concentration rises and approaches the critical free ATP-DnaA concentration $[D]_{\rm ATP,f}^\ast$, ATP-DnaA begins to bind cooperatively to the origin binding sites. This causes a weaker rise in the free DnaA concentration right before replication initiation as compared to a system without these origin binding sites (Fig. \ref{fig:fig_10_cascade}a and b). After an origin has been initiated, the origin binding sites become unavailable, causing an increase in the free DnaA concentration after replication initiation (Fig. \ref{fig:fig_10_cascade}a and b). While this increase should enhance the probability of other origins to fire replication as well, the weaker rise in the free concentration before replication initiation reduces the sharpness of the rise in the free ATP-DnaA concentration and should therefore lead to a decrease in the degree of synchrony. Indeed, comparing the simulations in which the origin-binding is modelled explicitly (Fig. \ref{fig:fig_10_cascade}c) to the previous model in which we simply used a Hill function for the opening probability (Fig. \ref{fig:fig_7}d) shows that the degree of synchrony is not significantly enhanced by the initiation cascade. The reason likely is that the positive effect of the initiation cascade is counterbalanced by the negative effect of a lower rise in the free concentration before replication initiation. 

\bibliography{Synchrony}

\end{document}